\documentclass[fleqn,usenatbib]{mnras}

\usepackage[T1]{fontenc}

\DeclareRobustCommand{\VAN}[3]{#2}
\let\VANthebibliography\thebibliography
\def\thebibliography{\DeclareRobustCommand{\VAN}[3]{##3}\VANthebibliography}

\newcommand\lal{Ly-$\alpha$ }
\newcommand\lab{Ly-$\beta$ }
\newcommand{\rev}[1]{{#1}}

\usepackage{graphicx}	% Including figure files
\usepackage{amsmath}	% Advanced maths commands
\usepackage{amssymb}	% Extra maths symbols

\title[Ly-$\alpha$ and \lab continuum reconstructions]{A comparison of quasar emission reconstruction techniques for $z\geq5.0$ Lyman-$\alpha$ and Lyman-$\beta$ transmission}

\author[S. E. I. Bosman et al.]{
Sarah E. I. Bosman,$^{1,2}$\thanks{E-mail: bosman@mpia.de}
Dominika \v{D}urov\v{c}\'{i}kov\'{a},$^{3}$ 
Frederick B. Davies,$^{4,2}$\newauthor
Anna-Christina Eilers$^{5}$
\\
$^{1}$Department of Physics and Astronomy, University College London, Gower Street, London WC1E 6BT, UK\\
$^{2}$Max-Planck-Institut f{\"u}r Astronomie, K{\"o}nigstuhl 17, D-69117 Heidelberg, Germany \\
$^{3}$New College, University of Oxford, Holywell Street, Oxford OX1 3BN, UK\\
$^{4}$Lawrence Berkeley National Laboratory, CA 94720-8139, USA\\
$^{5}$MIT Kavli Institute for Astrophysics and Space Research, 77 Massachusetts Ave., Cambridge, MA02139, USA
}

\date{Accepted XXX. Received YYY; in original form ZZZ}

\pubyear{2020}

\usepackage{newtxtext,newtxmath}

\begin{document}
\label{firstpage}
\pagerange{\pageref{firstpage}--\pageref{lastpage}}
\maketitle

\begin{abstract}
Reconstruction techniques for intrinsic quasar continua are crucial for the precision study of Lyman-$\alpha$ (Ly-$\alpha$) and Lyman-$\beta$ (Ly-$\beta$) \rev{transmission at $z>5.5$,} where the $\lambda<1215$\AA \ emission of quasars is nearly completely absorbed. While the number and quality of spectroscopic observations has become theoretically sufficient to quantify \lal transmission at $5.0<z<6.0$ to better than $1\%$, the biases and uncertainties arising from predicting the unabsorbed continuum are not known to the same level. In this paper, we systematically evaluate eight reconstruction techniques on a unified testing sample of $2.7<z<3.5$ quasars drawn from eBOSS. The methods include power-law extrapolation, stacking of neighbours, and six variants of Principal Component Analysis (PCA) using direct projection, fitting of components, or neural networks to perform weight mapping. We find that power-law reconstructions and the PCA with fewest components and smallest training sample display the largest biases in the \lal forest ($-9.58\%/+8.22\%$ respectively). Power-law extrapolations have larger scatters than previously assumed of $+13.1\%/-13.2\%$ over \lal and $+19.9\%/-20.1\%$ over \lab. We present two new PCAs which achieve the best current accuracies of $9\%$ for \lal and $17\%$ for \lab. We apply the eight techniques after accounting for wavelength-dependent biases and scatter to a sample $19$ quasars at $z>5.7$ with IR X-Shooter spectroscopy, obtaining well-characterised measurements for the mean flux transmission at $4.7<z<6.3$. %We compare the accuracy of reconstructions using optical-only spectra of $z>5.7$ quasars versus infrared spectra and estimate it would require $\sim80$ and $\sim50$ sightlines, respectively, to constrain \lal transmission within $5\%$ accounting for cosmic variance. 
Our results demonstrate the importance of testing and, when relevant, training, continuum reconstruction techniques in a systematic way.
\end{abstract}


%%%%%%%%%%%%%%%%%%%%%%%%%%%%%%%%%%%%%%%%%%%%%%%%%%

%%%%%%%%%%%%%%%%% BODY OF PAPER %%%%%%%%%%%%%%%%%%

\section{Introduction}

%Hydrogen reionisation is a crucial next step for cosmology
%importance of process: tied to first stars and galaxy formation
%its mid point from planck, future precision measurements at high (21cm) and low (lyman-a).

Hydrogen reionisation is thought to have been powered mostly by the first galaxies and quasars, which by redshift $z\sim 5.5$ were producing sufficient numbers of UV ionising photons to keep the intergalactic medium (IGM) near-completely ionised \citep{Robertson13}. The timing of reionisation, its progression, and its morphology on both small and large scales are therefore closely tied to the properties of the first sources (e.g.~\citealt{Robertson15, Madau15, Stark16, Dijkstra16}), making reionisation a crucial milestone for both cosmology and galaxy formation. 

While the \textit{Planck} satellite estimated the mid-point of reionisation \rev{at $z\simeq7.65\pm 0.73$} from the optical depth of electron scattering \rev{\citep{Planck20}} and upcoming $21$cm experiments promise constraints on the abundance of neutral hydrogen up to $z\lesssim20$ \citep{Yi08,Trott19}, the most accurate measurements currently come from measuring Lyman-$\alpha$ (Ly-$\alpha$) opacity towards $z>5.5$ quasars \citep{Fan00, Fan06, McGreer15} \rev{as lower limits}. \citet{GP} predicted that the \lal forest gives way to completely absorbed `Gunn-Peterson troughs' for IGM neutral fractions $\gtrsim 0.1\%$. Such features have now been detected down to $z\sim5.6$ \citep{Becker15}, but the transition from partial IGM transmission to full absorption is complex. 

On large scales, the coherence of \lal transmission on very large scales ($>160$ cMpc) and the unexpectedly large scatter between quasar sight-lines at the same redshift \citep{Becker15, Bosman18, Eilers18} have ruled out the simplest model of reionisation with a constant UV background (UVB) and inhomogeneities due only to density fluctuations \citep{Songaila04, Lidz06}. Current models are competing to match observations by incorporating further physical processes, such as a fluctuating mean free path of ionising photons \citep{Davies16,Daloisio18}, strong temperature fluctuations \citep{Daloisio15}, a contribution from rare sources \citep{Chardin15, Meiksin20}, or persisting neutral IGM patches \citep{Kulkarni19, Nasir20, Choudhury21}. The Lyman-$\beta$ (Ly-$\beta$) forest, which continues to display transmission after \lal has saturated, \rev{similarly reveals scatter in excess of predictions from density fluctuations alone} \citep{Oh05, Eilers19, Keating20}. 

In order to measure \lal and \lab transmission on large scales, it is crucial to accurately predict the intrinsic quasar emission before absorption by the IGM. The near-total absorption at $\lambda_\text{rest}<1215$\AA\ (the `blue side') means this requires extrapolation from the quasar rest-UV continuum at $\lambda>1220$\AA \ (the `red side'). In the past, this has been been done either via modelling of quasar emission by a power law (e.g.~\citealt{Bosman18}) or by training the reconstruction at $z<4$ via Principal Component Analysis (PCA) (e.g.~\citealt{Eilers18, Eilers20}). The accuracy of power-law reconstructions and PCA techniques is likely to become the leading source of uncertainty in \lal transmission studies at $z>5.5$ as the number of known quasars in the reionisation era will soon exceed $1000$ \citep{LSST, Brandt07} and deep spectroscopic observations are accumulating (D'Odorico in prep.).

Meanwhile, on small scales, many different statistical techniques are exploiting residual \lal transmission at $5.5<z<6.5$ where absorption troughs are punctuated by transmission spikes. As stronger IGM absorption makes it increasingly challenging to measure the \lal forest power spectrum directly at $z>5.2$ \rev{\citep{Nasir16, Onorbe17, Boera19}}, new techniques are instead using  the abundance and morphology of residual transmission spikes \citep{Chardin17-spikes, Gaikwad20}, the size distribution of absorption troughs \citep{Songaila02, Gallerani06, Gnedin17}, the fraction of absorbed pixels \citep{McGreer11, McGreer15} and correlations of the transmission with galaxies \citep{Becker18, Davies18-galaxies, Kakiichi18, Meyer19, Meyer20, Kashino20}. These methods are opening up novel ways to probe the temperature and ionisation state of the IGM and the sources responsible for reionisation, but now rely not only on the accuracy of continuum reconstruction methods but also, to varying extents, to the lack of wavelength-dependence of any residual biases arising during continuum reconstruction.

In this paper, we directly compare the performance of eight quasar reconstruction techniques on a common testing sample of quasars at $2.7<z<3.5$, where the intrinsic emission at $\lambda<1215$\AA \ can be reliably estimated. By carefully characterising the bias and uncertainties of these methods and their wavelength-dependence, we aim to reconcile current measurements of \lal opacity at $z>5.5$ and provide a grounding for reconstruction methods in the future. 

In Section \ref{sec:techniques}, we detail the continuum reconstruction techniques tested in this work and the related free parameters which include both empirical model-dependent reconstructions and machine-learning techniques. The selection of the test sample and our methods for testing and, when relevant, training the reconstruction methods are given in Section~\ref{sec:methods}. Section~\ref{sec:results} presents the results of the analysis in the form of mean biases and uncertainties for each technique and consequences for measurements of the mean \lal and \lab transmission at $z>5$. We summarise in Section~\ref{sec:ccl}.

Throughout the paper, we use a \textit{Planck} cosmology with $h=0.6732$, $\Omega_m=0.3158$ \citep{Planck18}. Distances are comoving and all wavelengths are given in the rest-frame of the emitting object unless otherwise specified.

\section{Continuum Reconstruction Techniques}\label{sec:techniques}

\begin{figure*}
\includegraphics[width=\textwidth]{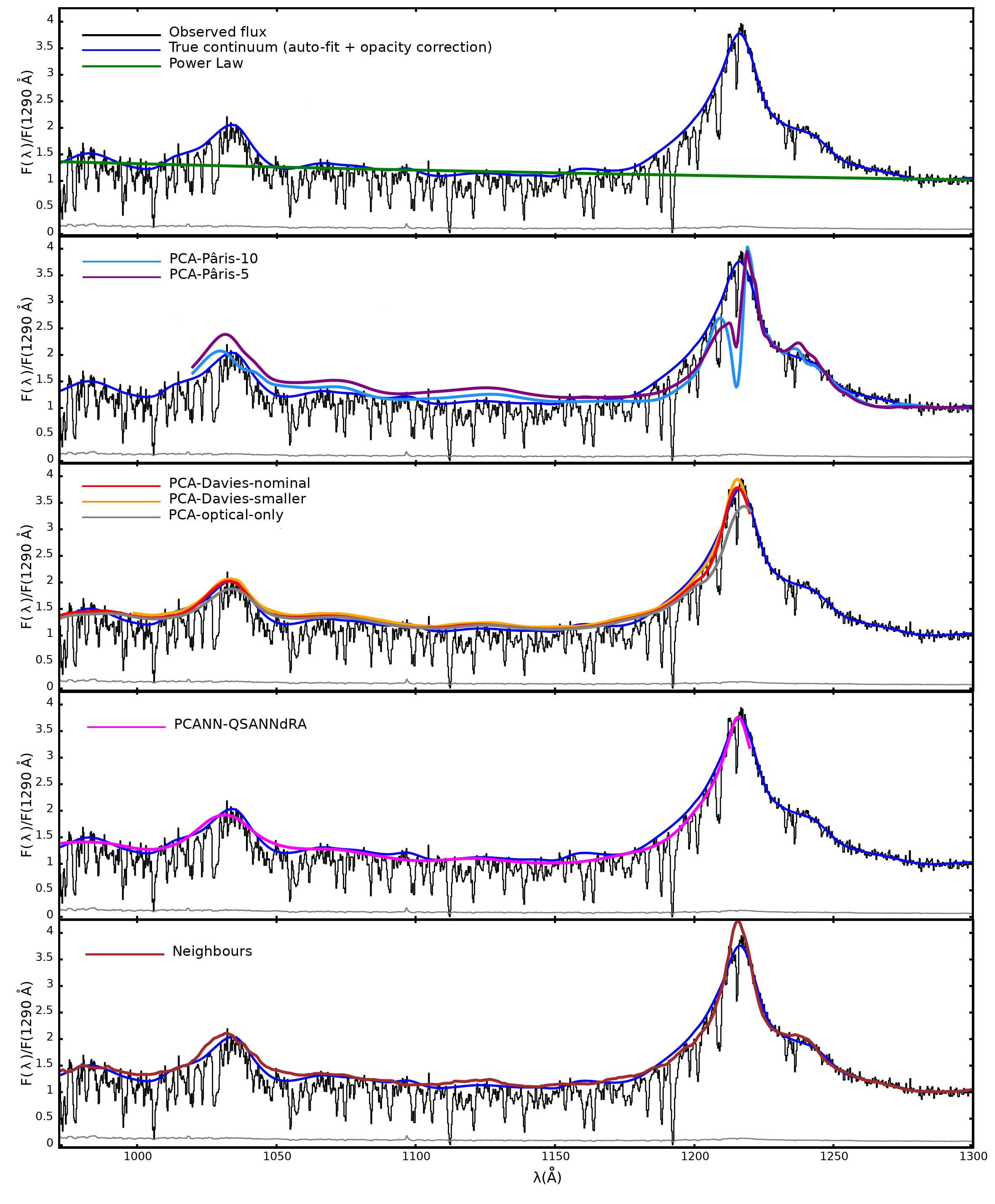}
\caption{Illustrative comparison of the eight quasar continuum reconstruction techniques on the same quasar, SDSS J131147.52+271826.7 at $z=2.974$. For each method, models are fit on the red side ($\lambda\geq1240$\AA) and used to make the predictions of the blue side ($\lambda\leq1290$\AA) shown in the five panels. The dark blue line shows the automatically-recovered continuum, used for testing the predictive accuracy of the models. The new PCAs were not trained to reproduce the shape of the \lal line.}
\label{fig:examp}
\end{figure*}

Quasar continuum reconstruction techniques used to analyse \lal transmission fall into three categories. First, explicitly model-dependent predictions include extrapolating a power-law shape of the spectral distribution function (SED) with best-fit parameters extracted from $\lambda>1220$ \AA \ (Section \ref{sec:PL}) and more complex models which derive empirical correlations between explicit quasar features, such as the shapes and strength of emission lines redwards and bluewards of the \lal line \citep{Greig17}. Second, the largest category is broadly model-independent machine learning techniques, usually in the form of PCA (Section~\ref{sec:pca}) but also including more refined Independent Component Analysis (ICA; as used in e.g.~\citealt{Rankine20}) and neural-network procedures (e.g.~\citealt{Fathivavsari20, Liu21}). 
Finally, one may use direct stacks of `nearest neighbours' to individual high-redshift quasars selected from populations of low-redshift quasars using varying (sometimes model-dependent) definitions of proximity (Section \ref{sec:NN}). Examples of the reconstruction techniques are shown in Figure~{\ref{fig:examp}} and more examples are available online \footnote{\url{http://www.sarahbosman.co.uk/research/supp20b}}.

\subsection{Power-law extrapolation} \label{sec:PL}

Quasar spectra display a wide range of narrow and broad emission lines with high ionisation energies $h \nu \geq 1$ Ry, which are thought to originate from photo-ionisation in accretion disks surrounding a supermassive black hole (SMBH) through processes involving synchrotron emission and Compton scattering of electrons \citep{Krolik88}. Underneath these emission lines, the mean SED of quasars assumes a comparatively simple power-law dependence, $F_\nu \propto \nu^{\alpha_\nu}$, over extended stretches of the optical wavelengths ($2000 < \lambda < 5000$ \AA), far UV (FUV; $2000<\lambda(\text{\AA})<1000$) and extreme UV (EUV; $\lambda<1000$\AA). Power-law extrapolations from the continuum redwards of \lal onto the bluewards side have been the most popular approach to reconstructing quasar intrinsic emission due to the method's relative simplicity \citep{Fan00, Fan06, McGreer15}. The uncertainties of power-law reconstruction have been assumed to be of order $\sim5\%$ or $\sim10\%$ (e.g.~\citealt{Fan02, Bosman18}).%, even though the location of potentially relevant breaks in the power-law SED shape are still under debate.

There is evidence for a break in the power-law form of quasar SEDs around the optical/FUV or FUV/EUV divisions. \citet{Richards01} analysed composite spectra from the Sloan Digital Sky Survey (SDSS, \citealt{SDSS}), finding slopes $\alpha_\nu \simeq -0.5$ over $1300<\lambda<5000$\AA. In contrast, studies of FUV and EUV quasars SEDs find much steeper slopes $\alpha_\nu \sim -1.7$ over $500<\lambda(\text{\AA})<1200$, with potential dependences of $\alpha_\nu$ on quasar luminosity and radio-loudness \citep{Telfer02, Scott04}. In a homogeneous analysis of the EUV, FUV and optical continua, \citet{Shull12} find roughly constant SED slopes over $2000<\lambda(\text{\AA})<5000$ of $\alpha_{\nu, \text{opt}} = -0.3$, which steepen significantly towards the FUV around $\lambda \sim2000$\AA \ to $\alpha_{\nu, \text{FUV}} = -1.4$ and again towards the EUV at $\lambda = 1000\pm 50$\AA \ to  $\alpha_{\nu, \text{EUV}} = -2.0$. If this is the case, extrapolations from the red side of \lal, $\lambda \gtrsim 1270$\AA, are expected to become increasingly biased at shorter wavelengths $\lambda<1000$\AA. Furthermore, power-law extrapolations do not capture the broad quasar lines superimposed on the continuum emission, which may give rise to strong wavelength-dependent biases. The characteristic velocity separations between broad lines give rise to spurious correlations on specific scales, which may be important to $z>5.5$ \lal studies in similar ways to $z<5.0$ (e.g.~\citealt{Lee12,MacDonald06,Onorbe17}).

In order to quantify the bias, we closely reproduce the procedure employed at $z>5.5$. The free parameters are the wavelength range being used, usually $1270\lesssim\lambda\lesssim 1450$ \AA, and the number of fitting iterations after masking of non-continuum pixels i.e.~`sigma-clipping'. Although this spectral range is less abundant in broad emission lines than longer wavelengths $\lambda>1450\AA$, it is nevertheless affected by the Si~{\small{II}} $1303$, O~{\small{I}} $1307$, C~{\small{II}} $1335$, Si~{\small{IV}} $1396$ and O~{\small{IV}}] $1404$ \AA \ broad lines (all of which are in fact doublets or triplets with separations $\gtrsim10$ times smaller than their typical widths). Furthermore, the broad emission lines show non-trivial velocity shifts with respect to the quasar's systemic redshift and to each other \citep{Greig17, Meyer19-qso} and the quasar redshifts estimated by the BOSS pipeline can be inaccurate by up to $\Delta z \simeq 0.05$ \citep{Hewett10,Coatman16}. It is therefore not feasible to fit a power-law continuum solely `in-between' the broad lines and iterative rejection criteria are employed instead.

In this work, we fit power-laws of the form $\text{PL}(\lambda) = k \lambda^{\alpha_\lambda}$ with $\alpha_\lambda = - (\alpha_\nu +2)$. We employ the full range $1270<\lambda(\text{\AA})<1450$ with three rounds of outlier rejection at $|F(\lambda) - \text{PL}(\lambda)| > 3 \sigma, 2.5\sigma, 2\sigma$. We find these exclusion criteria to be necessary and sufficient for fit convergence, and we illustrate the effect of changing the fitting wavelength range in the Appendix~\ref{app:PL}.

\begin{figure}
\includegraphics[width=\columnwidth]{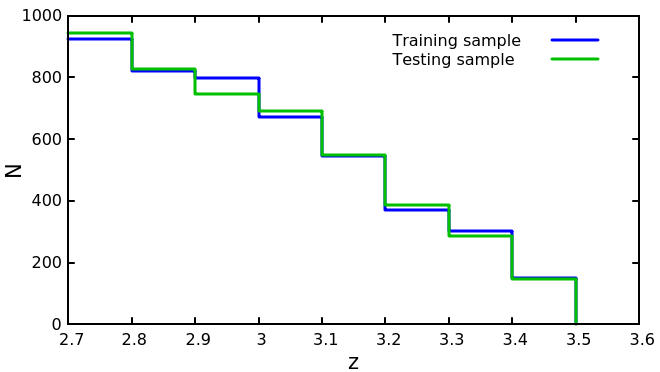}
\caption{Redshift distribution of the training and testing samples. Both samples contain $4597$ quasars and the objects were divided randomly.}
\label{fig:zhist}
\end{figure}
0

\subsection{Principal Component Analysis}\label{sec:pca}

Quasar spectra are a good target for PCA analyses due to their strong correlated spectral features: $\sim75\%$ of their observed optical/UV spectral properties can be captured with linear combinations of only $\sim 3$ components \citep{Francis92,Yip04,Suzuki05,Mcdonald05}. We outline here the principles of PCA decomposition and reconstruction, before giving more details on the specific PCAs used in our comparison in the following subsections. The steps involved in producing and applying a PCA reconstruction for the \lal transmission continuum are:\newline
\noindent \textbf{(1)} The identification of an appropriate training sample of N quasars for which the true underlying continuum $q(\lambda)$ at $\lambda<1220$ \AA \ (the blue side) and $\lambda>1220$\AA \ (the red side) can be accurately determined. The determination of $q(\lambda)$ can be manual or automated, and is usually different on the red side and the blue side (where the \lal forest requires special treatment) with a continuity requirement at the interface. \newline
\noindent \textbf{(2)} Decomposing both the red side continuum and the total (red and blue) continuum ($q(\lambda)$ into their principal components. Mathematically, this equates to computing the covariance matrix $\mathbb{C}$:
\begin{equation}
\mathbb{C}(i,j) = \frac{1}{N-1} \sum_{n=1}^N \left(q_{n}(\lambda_i) - \bar{F}(\lambda_i)\right)(q_{n}(\lambda_j) - \bar{F}(\lambda_j)),
\end{equation}
where $\bar{F}(\lambda)$ is the mean quasar flux over the entire training sample. One then finds the matrix $\mathbb{P}$ which diagonalises $\mathbb{C}$ i.e.~such that $\mathbb{C} = \mathbb{P}^{-1} \times \mathbf{\Lambda} \times \mathbb{P}$ where $\mathbf{\Lambda}$ is diagonal. The columns of $\mathbb{P}$ contain the eigenvectors/principal components $p_i(\lambda)$. \newline
\noindent \textbf{(3)} The mapping between the red side and the total spectrum is computed via the projection matrix $\mathbb{X}$. To do this, the weight matrices $\mathbb{W}$ on both sides are computed over all quasars in the training sample via
\begin{equation}
\mathbb{W}_{ij} = \int \left(q_i(\lambda) - \bar{F}(\lambda)\right) p_j(\lambda) d\lambda
\end{equation}
and finally the projection matrix is found in order to map between the sides: $\mathbb{W}_\text{red+blue} = \mathbb{X} \times \mathbb{W}_\text{red}$. At this stage, the number of principal components on both sides is customarily truncated to retain only those which account for most of the variance, although the criteria for defining  this cut-off vary widely.

\begin{figure}
\includegraphics[width=\columnwidth]{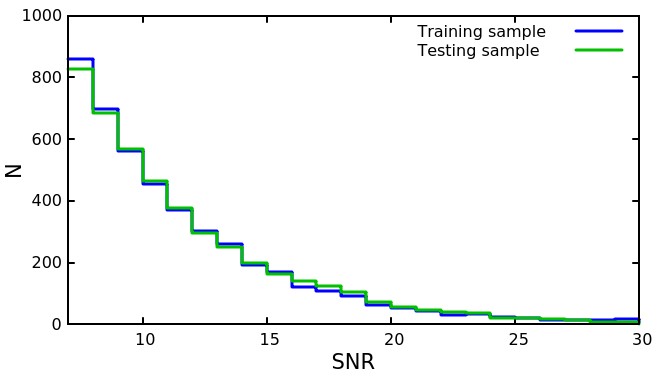}
\caption{SNR Distribution of the quasar spectra in the training and testing samples, measured at $1290\pm5$\AA. $\sim1\%$ of objects in both samples have SNR $>30$. Both samples contain $4597$ quasars and the objects were divided randomly.}
\label{fig:SNRhist}
\end{figure}

Armed with the mean flux $\bar{F}(\lambda)$, the two sets of principal components $p_i(\lambda)_\text{red}$ and $p_i(\lambda)_\text{red+blue}$ and the projection matrix $\mathbb{X}$, one can now make a prediction of the blue continuum for any new quasar outside the training set as follows:\newline
\noindent \textbf{(4)} Extract the red-side continuum $q_r(\lambda)$ of the new quasar with the same technique as used in step (1).\newline
\noindent \textbf{(5)} Project or fit the set of components weights $w_{i, r}$ such that $q_\text{red}(\lambda) - \bar{F}(\lambda) =  \sum_i w_{i, r} p_{i, \text{red}}(\lambda)$. In theory, this can be achieved cleanly via projection of $q_\text{red}(\lambda)$ in the basis of the $p_{i, \text{red}}(\lambda)$:
\begin{equation}
w_{i, r} = \int \left(q_\text{red}(\lambda) - \bar{F}(\lambda)\right) p_{i, \text{red}}(\lambda) d\lambda;
\end{equation}
but in practice, the truncation of the principal components at step (3) and the desire to obtain uncertainties on the weights $ w_{i, r}$ means that direct fitting in wavelength space is very frequently used instead. \newline
\noindent \textbf{(6)} The vector of weights on the red+blue sides, $\mathbf{w_{r+b}}$, is obtained from the weights on the red side, $\mathbf{w_r}$, by multiplying with the projection matrix, $\mathbf{w_{r+b}} = \mathbb{X} \mathbf{w}_r$. The continuum prediction is given by $\text{Pred}(\lambda) = \bar{F}(\lambda) + \sum_i w_{i, r+b} p_{i, \text{red+blue}}(\lambda)$.

Although important differences between PCA methods occur at stages (1), (3) and (5) as will be discussed below, the most crucial difference is in the choice of training sample in step (1). 
Quasars at $z>6$ have shown signs of intrinsic evolution compared to $z<5$, displaying extreme velocity shifts of the highly-ionised broad emission lines more frequently at early times at equal quasar luminosity \citep{Shen08, Richards11,Mazzucchelli17, Meyer19-qso, Shen19}. It is therefore extremely important that the training sample for PCA be large enough to capture the relatively rare analogue objects at $z<5$. If extreme broad line shifts are caused by a distinct physical process, it is also crucial that sufficient principal components are retained at step (3) to adequately describe this population. Finally, one should be wary that the continuum fitting at steps (1) and (4) be not biased against quasar spectra with strong line shifts, weak emission lines, or lower signal to noise ratios (SNR). PCA is trained directly to reproduce the intrinsic continua recovered in step (1). For example, the presence of strong damped \lal absorbers (DLAs) which are not correctly excluded from the training continuum fitting will introduce random  features which PCA cannot match and reduce its predictive power. To illustrate the cumulative importance these effects, we test two models which differ solely in step (1): \textit{PCA-Davies-smaller} and \textit{PCA-Davies-nominal} (Section~\ref{sec:Dnew}).

\subsubsection{Classical quasar PCA}

The first PCA reconstruction we test is through the components of \citet{Paris11} (P11), which have previously been used to study \lal transmission at high-$z$ \citep{Eilers17, Eilers18}. P11 provide $10$ principal component vectors for both the $1020<\lambda(\text{\AA})<2000$ (total) range and the $1216 < \lambda < 2000$\AA \ (red) range, as well as a projection matrix for linking the two. Their training sample consists of 78 quasars at $2.82<z<3.00$ from SDSS-DR7. The quasars were selected to have SNR $>14$ over the red-side continuum, and the sample was manually cleaned to avoid any broad absorption line (BAL) quasars, any DLAs, and any spectra with reduction issues. The continua were fit with a spline manually. P11 estimated the uncertainty in predicting the \lal forest continuum via the `leave one out' method, i.e.~training on 77 quasars and evaluating the prediction on the 78$^\text{th}$. P11 report uncertainties of $9.5\%$ at the 90$^\text{th}$ percentile. To test the prediction on a larger collection of quasars, we automate the continuum spline fitting while retaining the ban on BAL quasars (see Section~\ref{sec:cont-fit}). Following P11, we use all $10$ components and conduct weight determination (step (5) above) via re-binning and direct projection. We label this method \textit{PCA-P\^aris-10} in the rest of the paper.

\begin{table}
\begin{tabular}{l l l l}
Quasar name & $z_\text{sys}$ & SNR & Ref. \\
\hline
J1120+0641 & 7.085 & 48 & \citet{Mortlock11} \\
PSOJ011+09 & 6.4693 & 17 & \citet{Mazzucchelli17} \\
J0100+2802 & 6.3258 & 119 & \citet{Wu15} \\
J1030+0524 & 6.3000 & 35 & \citet{Fan01} \\
J0330-4025 & 6.239 & 13 & \citet{Reed17} \\
PSOJ359-06 & 6.1718 & 41 & \citet{Wang16} \\
J2229+1457 & 6.1517 & 11 & \citet{Willott10} \\
J1319+0950 & 6.1333 & 104 & \citet{Mortlock09} \\
J1509-1749 & 6.1225 & 66 & \citet{Willott07} \\ 
PSOJ239-07 & 6.1098 & 36 & \citet{Banados16} \\
J2100-1715 & 6.0806 & 30 & \citet{Willott10} \\
PSOJ158-14 & 6.0681 & 38 & \citet{Chehade18} \\
J1306+0356 & 6.0332 & 77 & \citet{Fan01} \\
J0818+1722 & 6.00 & 135 & \citet{Fan06} \\
PSOJ056-16 & 5.967 & 54 & \citet{Banados16} \\
J0148+0600 & 5.923 & 139 &\citet{Jiang15} \\
PSOJ004+17 & 5.8165 & 27 & \citet{Banados16} \\
J0836+0054 & 5.810 & 85 & \citet{Fan01}\\ 
J0927+2001 & 5.772 & 93 & \citet{Fan06} \\
\hline
\hline
\end{tabular}
\caption{Quasars included in the computation of the mean \lal and \lab transmitted flux at $z>5$, selected to have IR X-Shooter spectra with BOSS-equivalent SNR $>7$.}
\label{tab:quasars}
\end{table}

Using the P11 PCA components to reconstruct the continua of $z>5.5$ quasars, \citet{Eilers17} noticed that restricting the fit to only the first $7$ or $5$ PCA components gave a qualitatively better fit to the red side in every case for their sample of $34$ quasars. One possible explanation is the presence of redshift errors in the P11 training sample, which components $6$ to $10$ seem predominantly designed to correct. To adjust to this difference, \citet{Eilers17} opt to use the first $3$, $5$ or $7$ components of the P11 PCA depending on the qualitative appearance of the red-side fit, and use the PCA components from \citet{Suzuki06} instead on $7$ quasars where none of those options were satisfactory. This more careful process is difficult to automate, so we choose to only test the $5$-component fit which was used most commonly ($\sim59\%$) for their sample. Following \citet{Eilers17}, we conduct weight determination via $\chi^2$ fitting and label this technique \textit{PCA-P\^aris-5}. We indeed find the fits with only $5$ components to be qualitatively different from fits using all $10$ (see e.g.~Figure~\ref{fig:examp}). It is important to note the PCA's projection matrix is not re-trained when changing the number of components, but only truncated. %Since increasing the number of components always formally decreased the best-fit reduced $\chi^2$, 

\subsubsection{New quasar PCA}\label{sec:Dnew}

The SDSS-III Baryon Oscillation Spectroscopic Survey (BOSS) and the SDSS-IV Extended BOSS (eBOSS) obtained low resolution ($R\sim2000$) spectra of the \lal forest of over $290,000$ quasars at $2<z<5$ \citep{BOSS, eBOSS}. \citet{Davies18-PCA} (D18) leveraged the SDSS-DR12 BOSS quasar catalogue \citep{Paris17} to create new quasar PCA decompositions, following the steps described above with a few modifications. D18's original training sample includes $12,764$ quasars with $2.09<z<2.51$ and SNR $>7.0$ and is relatively devoid of BALs and wrong quasar identifications by virtue of the visual inspection flags in the SDSS DR12 quasar catalogue. New additional BAL exclusion methods were also introduced. The red and blue sides of the continuum are split entirely into $\lambda<1220$\AA \ and $\lambda>1220$\AA \ instead of predicting the overall continuum in step (2). The method implements a continuity condition at the interface, and fitting is performed with an automated spline fitter similar to the one in this paper but without explicit DLA masking (see Section~\ref{sec:cont-fit}). Two major additions to the procedure in Section~\ref{sec:pca} are: \begin{itemize}
\item The PCA-building and fitting procedures are conducted in flux log space, using $\text{log}(F(\lambda))$ instead of $F(\lambda)$. This enables the power-law component of the quasar continuum to be more naturally  represented by an additive component, rather than a multiplicative one.
\item At the fitting stage, a small shift in redshift ($|\delta z|<0.05$) is fit at the same time as the PCA component weights. Consequently, weight determination is always conducted via $\chi^2$ minimisation.
\end{itemize} The original PCA in D18 was aimed at predicting the shape of the \lal emission line rather than the bluewards continuum \citep{Davies18-DW,Wang20}. Prediction errors at $\lambda\sim 1210$ \AA \ were estimated to be around $6-12\%$. To predict the \lal forest continuum, we shift the dividing point between the red and blue sides from $\lambda=1280$\AA \ to $\lambda=1220$\AA \ in order to access more emission line properties.

We test three versions of the PCA generated using D18's procedure, all using $10$ red-side components and $6$ blue-side components. The first version is trained on BOSS DR12, employing $2352$ spectra with SNR $>10$ at $2.65<z<3.00$ covering the range $972<\lambda(\text{\AA})<2500$. The intrinsic continuum fitting is the same as in D18 and the fluxing correction described in \citet{Margala16} has been applied to address a newly-known issue with blue quasar slopes in BOSS DR12. We label this method \textit{PCA-Davies-smaller}.

Secondly, we train a new PCA following D18's methodology but using our common quasar training set with $4579$ quasars based on the larger eBOSS-SDSS DR14 catalogue (see Section~\ref{sec:sample}). Crucially, this PCA is trained using the same continuum recovery algorithm on the blue side as we use for testing the PCA's predictive performance. The contrast between these two versions illustrates the variance in PCA predictive performance arising purely from sample size, sample purity, and the recovery of the blue-side continuum. We label this method \textit{PCA-Davies-nominal}.

Thirdly, we test a PCA identical to the previous one in all points, but using a restricted wavelength range of $972<\lambda(\text{\AA})<1450$. Spectra of $z>5.7$ quasars are frequently only available in the observed-optical wavelengths, in which case only a power-law reconstruction can be used. We therefore wish to test whether PCA can outperform power-law reconstructions under this constraints. We label this method \textit{PCA-optical only}. 

\subsubsection{Neural-network-mapped quasar PCA}

The final version of PCA we implement is the neural-network-mapped quasar PCA-fitting algorithm \textit{PCANN-QSANNdRA} described in \citet{Durovcikova20}. Like the previous technique, \textit{PCANN-QSANNdRA} was originally designed to predict the shape of the \lal emission line rather than the bluewards continuum. The training sample included $13,703$ quasars from eBOSS DR14 over $2.09<z<2.51$ and SNR $>7$ \citep{Paris18}. The continuum was smoothed using a custom routine which employed a random forest to tag intervening absorbers and exclude quasars with strong DLAs. The red side ($1290<\lambda(\text{\AA})<2900$) and the blue side ($1191.5<\lambda(\text{\AA})<1290$) of the continuum are described by $63$ and $36$ PCA components, respectively, with the number of components chosen to contain $99\%$ of the variance in the training sample. The red and blue sides are completely separated, as in D18.

Instead of the weights of the PCA component vectors being mapped linearly following Equation (2), the translation between the blue and red side weights is performed by a four-layered fully connected neutral network. 
This method introduces an extra step of `double standardisation' whereby the mean across the training sample is subtracted at each wavelength before and after PCA composition, and divided by the error array. 
The prediction uncertainties were calculated by a committee of $100$ independently-trained neural networks. No simultaneous fitting of a shift in redshift is performed. Owing to the freedom to enact non-linear mapping of PCA coefficients, \textit{PCANN-QSANNdRA} slightly outperformed D18's PCA method with uncertainties of $\sim 6-10\%$ at $\lambda\sim 1210$\AA.  

%\vskip2cm
In this paper, we use an identical methodology to train a neural-network-mapped PCA to reproduce the continuum bluewards of \lal using the eBOSS DR14 common training and testing samples described in Section~\ref{sec:sample}. 
\rev{
As in \citet{Durovcikova20}, we use a $50:50$ split for training and validation (both within the eBOSS training sample) and adapt the network to predict the Lyman-series continuum using an identical procedure. In order to capture $99\%$ of the variance, we find that $53$ PCA components are needed for the red side and $42$ for the blue side. The optimal architecture in \citet{Durovcikova20} consisted of four layers with $53-40-40-42$ neurons each, a batch size of $500$ (corresponding to the number of training samples passed through the network before the weights get updated), and $80$ training epochs. To optimise training, we first adjusted the loss function so that prediction errors around $\lambda\sim1040$\AA \ and $\lambda\sim1215$\AA \ are weighed more heavily to anchor the prediction where the scatter is largest. The ability to adjust training loss function weights in this manner makes neural-network-mapped quasar PCAs more flexible than standard PCA. We then performed a four-dimensional grid search over the number of neurons in the middle two layers, the batch size, and number of training epochs. The optimal architecture consisted of layers with $53-45-45-42$ neurons, batch size of $500$, and $80$ training epochs. Compared to the original architecture, the optimal one reduced the mean bias below $1\%$ (potentially driven mainly by the updated loss function) but had almost no impact on reducing the prediction error ($<0.2\%$ difference). More details of the neural network architecture and training procedure are given in \citet{Durovcikova20}, section 2.3.
}
%The neural-network architecture has four layers with $53-40-40-42$ neurons each. To contain $99\%$ of the variance, $53$ components are needed for the red side and $42$ for the blue side. The pivot between the red and blue sides occurs at $\lambda = 1290$\AA. In order to optimise the training for predictions of the Lyman-series continuum, the loss function was adjusted so that prediction errors around $\lambda\sim1040$\AA \ and $\lambda\sim1215$\AA \ are weighed more heavily to anchor the prediction where the scatter is largest. The ability to adjust training loss function weights in this manner makes neural-network-mapped quasar PCAs more flexible than standard PCA.

\subsection{Nearest Neighbours} \label{sec:NN}

Stacks of $z<3.0$ quasar spectra are often used to predict the intrinsic emission of $z>5$ quasars and capture features such as broad emission lines in a model-independent way (e.g.~\citealt{Berk01,Cool06,Simcoe11}). Stacks specifically tailored from `neighbours' which match specific features of the red side, such as strong blueshift of the broad C~{\small{IV}} emission line, have been shown to perform better than blind stacks \citep{Mortlock11, Bosman15} since those features are known to correlate with properties on the blue side \citep{Baldwin77,Richards11,Greig17}. 

We systematically test the performance of stacks of nearest neighbours for predicting intrinsic emission. To select neighbours, we first fit all quasars with continua on the red side using an automatic spline fitter 
\rev{with fixed point spacing of $\sim1000$km s$^{-1}$,} rejecting metal absorbers as described in \ref{sec:cont-fit}. \rev{In the absence of \lal forest absorption, we find this point spacing better captures the shape of the emission lines (a point spacing of $\sim1700$km s$^{-1}$ was used to fit the blue side).} 
For each quasar $q_i$ in the testing sample, the Euclidean distance $d_{i,j}$ to all quasars in the training sample $q_j$ is then computed over the red side, $1220<\lambda(\text{\AA})<2000$, 
\begin{equation}
d_{i,j} = \sqrt{  \sum\limits_{\lambda>1220\text{\AA}}\left( q_i(\lambda) - q_j(\lambda) \right)^2 },
\end{equation}
in order to identify the $40$ quasars from the training sample with the smallest $d_{i,j}$. The known blue-side continua of the neighbours are then interpolated onto a common wavelength array and averaged. The distance between spectra could be defined in different ways, e.g.~weighting more heavily emission lines known to correlate with blue side properties, or calculating the distance after subtracting a power-law fit first. These differences will be most important when the number of neighbours of a specific quasar is limited, and we leave a detailed exploration to future work.
%Can be defined in many ways. Euclidean distance isn't a good metric, but it works really well and it's not clear (to me) why. Neural networks are confusing. Write this paragraph. (e.g.~\citep{Mortlock11, Bosman15})

\begin{table} 
\begin{tabular}{c l l l l}
$z$ & $z_\text{mean}$ & $\int dz$ & $\langle F \rangle_\text{\textit{PCA-nominal}}$ & $\langle F \rangle_\text{\textit{QSANNdRA}}$ \\
\hline
\lal & & & \\
$4.7 - 4.9$ & $4.826$ & $0.489$ & $0.1856_{-0.0132}^{+0.0126}$& $0.1974_{-0.0144}^{+0.0140}$\\
$4.9 - 5.1$ & $5.015$ & $1.826$ & $0.1258_{-0.0046}^{+0.0045}$& $0.1330_{-0.0048}^{+0.0049}$\\
$5.1 - 5.3$ & $5.201$ & $2.848$ & $0.0991_{-0.0031}^{+0.0031}$& $0.0979_{-0.0032}^{+0.0032}$\\
$5.3 - 5.5$ & $5.399$ & $3.194$ & $0.0765_{-0.0025}^{+0.0024}$& $0.0742_{-0.0026}^{+0.0026}$\\
$5.5 - 5.7$ & $5.596$ & $2.804$ & $0.0435_{-0.0021}^{+0.0021}$& $0.0432_{-0.0022}^{+0.0022}$\\
$5.7 - 5.9$ & $5.789$ & $2.022$ & $0.0236_{-0.0024}^{+0.0024}$& $0.0222_{-0.0025}^{+0.0024}$\\
$5.9 - 6.1$ & $5.981$ & $0.778$ & $0.0115_{-0.0036}^{+0.0036}$& $0.0117_{-0.0036}^{+0.0036}$\\
$6.1 - 6.3$ & $6.188$ & $0.290$ & $<0.0131$ & $<0.0133$\\
\hline
\hline
\lab & & & \\
$5.5 - 5.7$ & $5.605$ & $0.719$ & $0.0534_{-0.0081}^{+0.0080}$& $0.0601_{-0.0076}^{+0.0086}$\\
$5.7 - 5.9$ & $5.801$ & $1.321$ & $0.0358_{-0.0042}^{+0.0042}$& $0.0425_{-0.0042}^{+0.0046}$\\
$5.9 - 6.1$ & $5.979$ & $0.732$ & $0.0181_{-0.0029}^{+0.0029}$& $0.0179_{-0.0028}^{+0.0031}$\\
\hline
\hline
\end{tabular}
\caption{Redshift intervals, mean redshifts, and effective redshift length probed by the sample of $z>5.7$ quasars. We only report redshift bins with $\int dz>1$. Limits are given at the $2\sigma$ level.}
\label{tab:dz}
\end{table}

\section{Methods}\label{sec:methods}

We aim to test the quasar continuum predictions described in Section~\ref{sec:techniques} in a consistent way and on a common sample. Each method is applied in a similar way as previous work, and the PCA methods requiring training samples (\textit{PCANN-QSANNdRA} and \textit{PCA-Davies-nominal}) are trained on the same sample. Section~\ref{sec:sample} describes the selection of the training and testing samples from the eBOSS DR14 catalogue. We then compare predictions to an automatic continuum-tracer bluewards of the \lal emission line. Because we predict the Ly-$\beta$ continuum down to $\lambda = 972$\AA, we need to use quasars with $z>2.7$ for which the Ly-$\alpha$+\lab forest absorption is dense. We use an automated spline fitter for the \lal forest with masking of individual absorbers, which we describe in Section~\ref{sec:cont-fit}. The estimation of the methods' biases and uncertainties is described in Section~\ref{sec:bias} and their application to $z>5.5$ \lal and \lab opacities in Section~\ref{sec:highz}.

\subsection{Training and testing sample selection} \label{sec:sample}

The selection of the training and testing samples is conducted from the eBOSS DR14 quasar catalogue using the \texttt{igmspec} module of the \texttt{specdb} interface \citep{igmspec}. We use quasars in the redshift range $2.7<z<3.5$ to ensure coverage of the \lab forest in the BOSS camera spectral \rev{observed wavelength} range of $3600<\lambda_{\text{obs}}(\text{\AA})<10400$ \citep{Smee13}, while IGM absorption becomes too strong beyond $z>3.5$ to enable secure recovery of the true continuum.

We impose a cut of SNR $>7$ at $\lambda=1290$\AA \ to ensure that both the red and blue side can be fit with a spline continuum even in the presence of a high spectral slope. The eBOSS catalogue contains $8118$ objects with those criteria according to the automated pipeline. However, we perform our own calculation of SNR over the $1285<\lambda(\text{\AA})<1295$\AA\ spectral window after masking sky-lines and identify $12,597$ objects instead. 

As opposed to the BOSS DR12 catalogue, eBOSS DR14 was not inspected visually and lacks specific data quality flags for BAL quasars. We conduct an automated search for BALs as follows. First, we fit each spectrum with an automated spline with initial fixed points in $\sim 7.5\AA$ \ intervals, corresponding to $1700$ km s$^{-1}$. We use the fitting procedure developed by \citet{Young79} and \citet{Carswell82} as implemented by \citet{Dallaglio08}, without masking of absorbers. We then normalise the spectra based on the value of the continuum at $\lambda = 1290$\AA, $F(1290)$. Finally, we cut from the sample any quasars whose automated fit drops below $F/F(1290) = 0.2$ across 
$1220<\lambda(\text{\AA})<1950$ or below $F/F(1290) = 0.7$ over the $1290<\lambda(\text{\AA})<1570$ window, which would correspond to a C~{\small{IV}} BAL ($1164$ objects or $9\%$ of the sample). Using the same wide-spline fit, we identify strong absorption features which would impair the continuum recovery on the blue side as objects whose continuum drops below $F/F(1290) = 0.5$ over $972<\lambda(\text{\AA})<1220$ ($2104$ objects or $17\%$ of the sample). This occurrence rate is a factor $\sim3$ higher than expected for DLAs alone at $z\sim3$ (e.g.~\citealt{Prochaska09, Crighton15}), suggesting we are also excluding weaker Lyman-limit systems. We further exclude any quasar with missing flux information over $\geq7$ consecutive spectral pixels ($34$ objects).

\begin{figure*}
\includegraphics[width=\textwidth]{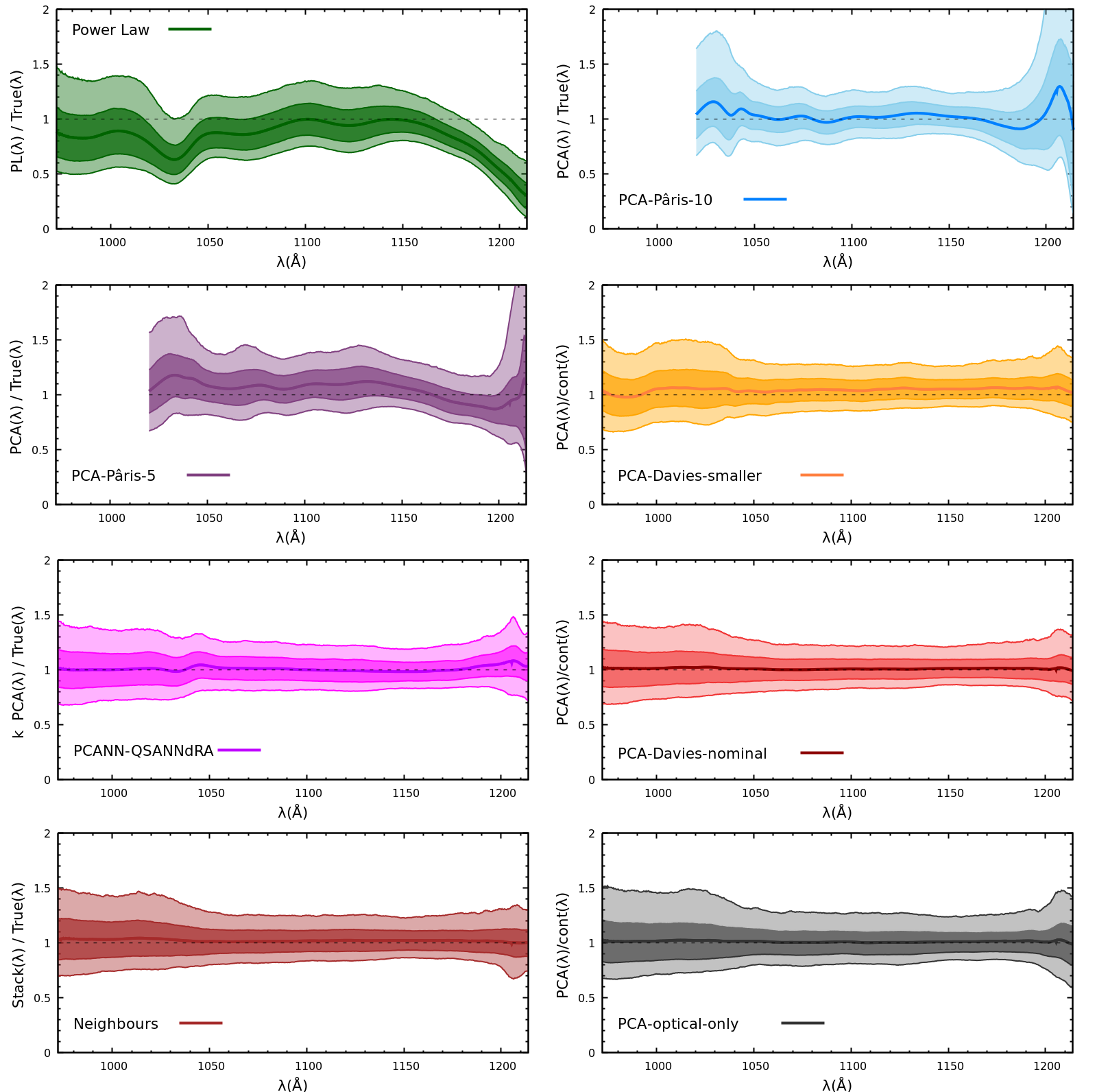}
\caption{The mean error (bias) in continuum prediction for eight quasar emission reconstruction methods, evaluated over the testing sample of $4579$ quasars at $z<3.5$. The shaded regions correspond to two-sided 1 and 2$\sigma$ contours containing 68\% and 95\% of the scatter in prediction error. The \textit{Power-Law}, \textit{PCA-P\^aris-10}, \textit{PCA-P\^aris-5}, \textit{Neighbours} predictions are compared to the true continuum, i.e.~the automated continuum corrected for IGM absorption at $z\sim2.6$ by a multiplicative constant. The \textit{PCA-Davies} models are compared to the continuum recovery on which they were trained. The \textit{PCANN-QSANNdRA} method is compared to the true continuum, divided by a constant $k$ accounting for the different IGM absorption biases of automatic continuum recovery.}
\label{fig:compar}
\end{figure*}

Finally, we perform visual inspection of the remaining quasars to exclude incorrect pipeline identification or redshifts. We identify $114$ objects with redshift errors (usually due to mis-identification of the Mg~{\small{II}} line as Ly-$\alpha$), $18$ quasars with proximate DLAs too strong to use and $5$ extra BAL quasars not caught by our automated procedure due to the absorption having width $\Delta v<1000$ km s$^{-1}$. The final sample contains $9158$ usable quasars, that are divided randomly and evenly into a testing sample of $4597$ objects and a training sample of $4597$ objects. The redshift and SNR distributions of the samples are shown in Figure~\ref{fig:zhist} and Figure~\ref{fig:SNRhist}, respectively.

\begin{figure*}
\includegraphics[width=\textwidth]{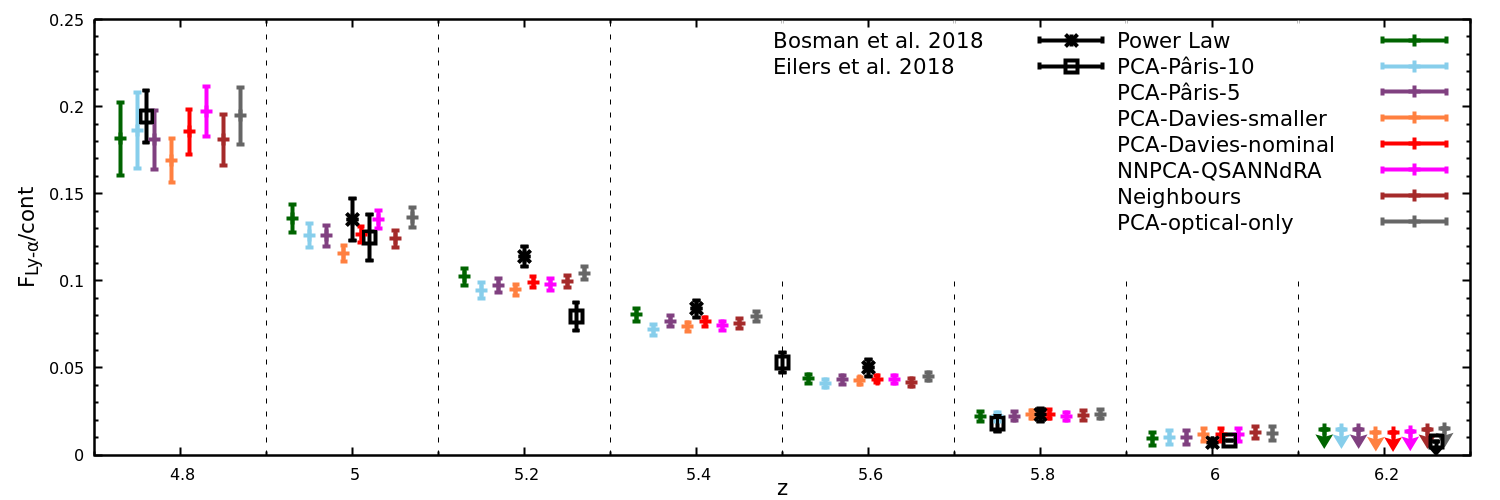}
\caption{Mean transmitted \lal flux along the sightlines of quasars at $z>5.7$. The coloured points correspond to predictions made from eight reconstruction methods, with bias corrections applied from testing on the $2.7<z<3.5$ eBOSS testing sample. Each set of coloured points correspond to stack in redshift between the dotted lines. The uncertainties account for both the $1\sigma$ uncertainties in reconstruction accuracy and observational errors. All techniques agree within $1.5\sigma$. The measurements from \citet{Eilers18} and \citet{Bosman18} use different samples and a different binning of $\Delta z = 0.25$ and $0.2$, respectively.}
\label{fig:lya}
\end{figure*}

\subsection{Continuum fitting} \label{sec:cont-fit}

We use the following procedure for recovering the underlying quasar continuum on the blue side. 
Quasar emission lines in the \lal and \lab forest: N~{\small{III}} 990 \AA, O~{\small{VI}} 1033\AA, N~{\small{II}} 1085\AA, Fe~{\small{III}} 1125\AA \ and C~{\small{III}} 1175\AA, are more widely spaced and broader on average than emission lines on the red side \citep{Shull12}. We therefore start with fitting a spline to the spectrum with initial fixed points $\sim 1700$ km s$^{-1}$ intervals as above (more specifically, $25$ pixels in the wavelength array). The spline-fitting procedure of \citet{Dallaglio08} iteratively excludes individual pixels within the bins which have the largest negative deviations from the fit via asymmetric sigma clipping. Convergence is achieved when the standard deviation of \rev{the} retained flux pixels is below the average observed noise in each bin.

However, we find that the fit has a tendency to return unphysical ``wiggles'' when it interpolates over absorbers with $\Delta v \gtrsim 0.5$ of the \rev{distance between fixed points due to the procedure} being stuck to its `first guess' even after all the absorbed pixels are masked. To circumvent this, we implement a more drastic absorber-masking procedure and run the fitter again after explicitly removing all information on pixels identified as absorbers after a first pass. Additionally, we use a stringent requirement that $>40\%$ of all pixels in a bin must be un-masked in order for that bin to factor into the fitting at all. We find the resulting continuum recovery to be far better at interpolating across strong H~{\small{I}} absorbers, yet nearly identical to the old procedure in their absence (Appendix~\ref{app:contfit}).

\rev{The automatic continuum fitting introduces a non-zero intrinsic scatter on the recovered `true' continua arising from stochasticity in the fit quality. \citet{Dallaglio09} estimate this scatter to be $\sim2.5\%$ at $z=3$. The authors note that the procedure also introduces wavelength-dependent biases of order $2-3\%$, mostly by occasionally under-estimating the continuum over the O~{\small{VI}} line when its peak coincides with \lal absorption. We expect the improvements made to the method, as described above, should reduce these biases. The extra scatter from the automatic continuum fitting is sub-dominant compared to the measured continuum prediction scatter (Section \ref{sec:results}) and we do not consider it further. Wavelength-dependent biases in the procedure are potentially an issue, especially considering that four of the reconstruction methods used different procedures to determine the `true' continua (\textit{PCA-P\^aris-10}, \textit{PCA-P\^aris-5}, \textit{PCA-Davies-smaller} and \textit{PCANN-QSANNdRA}). We nevertheless neglect this effect, since: (1) The wavelength-dependent biases of the four other reconstructions methods are not known, and two of them involved manual by-eye fitting which is impossible to reproduce formally; (2) Any extra wavelength-dependent biases did not stop \textit{PCANN-QSANNdRA} from achieveing the lowest mean bias of any techniques (Section \ref{sec:results}); (3) A full forward-modelling of these effects on simulated spectra, following \citet{Dallaglio09}, is beyond the scope of this work.}

\subsection{IGM correction at $2.0<z<3.0$}

The \lal forest is unresolved at the resolution of eBOSS, which makes automatic continuum fits liable to be biased \citep{Dallaglio09}. To quantify this, we measure the total transmitted flux  at $2.0<z<3.0$ in our testing and training samples and compare to fiducial literature values normalised to higher resolution spectra \citep{Faucher08,Becker13}. Our samples have a \lal forest midpoint of $z=2.634$. We measure a mean $\langle F/\text{fit}\rangle = 0.8152 \pm 0.0004$ over $2.6<z<2.7$, compared to $\langle F/\text{cont}\rangle = 0.7545\pm0.0088$ in \citet{Becker13}. The correction is therefore $\text{fit}/\text{cont} = 0.9255$. This is agreement with \citet{Dallaglio09}, who measured this bias for the original continuum fit we use here, before the tougher DLA exclusion described in the previous section. We note that in order to test techniques over a large wavelength range, we excluded sightlines containing DLAs entirely instead of masking them when measuring the mean flux. Since the presence of DLAs correlates with increased sightline opacity \citealt{Perez18}, our correction will in principle be biased to be slightly too small. This is a higher level effect which we neglect.

The evolution of IGM absorption over $2.0<z<3.0$ adds scatter to continuum predictions around the redshift mid-point. A more refined approach would be to apply a redshift-dependent IGM correction to the \lal forest of each individual quasar before training a PCA. To evaluate how much such a procedure would reduce the scatter, we measure the IGM correction at redshifts $z=2.42$ and $z=2.83$ which encompass $86\%$ of the \lal forest transmission in our sample. The IGM corrections are $\text{fit}/\text{cont} = 0.9530$ and  $\text{fit}/\text{cont} = 0.8978$, respectively. This corresponds to less redshift evolution than found by \citet{Dallaglio09}, perhaps owing to our refinements. We therefore find the `intrinsic' scatter due to IGM absorption evolution across our samples is of the order $\pm3\%$, much below the measured $1\sigma$ continuum prediction scatter of $\pm9\%$ which we will show in Section \ref{sec:results}. Including the correction would only lower the scatter to $\sim8\%$, and we are therefore satisfied with applying a single correction corresponding to the sample's mid-point.

\citet{Paris11} independently calibrated the continuum fits on which their PCA is trained manually, and find no residual compared to measurements of mean flux transmission at $z\sim3$. We will therefore test the \textit{PCA-P\^aris-10}, \textit{PCA-P\^aris-5}, \textit{Power-Law} and \textit{Neighbours} methods against the true continuum, i.e.~the corrected continuum fits. The three versions of \textit{PCA-Davies} were trained on the continuum fits themselves, and are therefore directly comparable to them. We note that re-training the PCAs on continua rescaled by a constant will not make a difference, since the overall normalisation is already an independent parameter. Finally, the \lal forest continuum-fitting method of the \textit{PCANN-QSANNdRA} technique could, \textit{a priori}, have a different bias due to unresolved IGM absorption. We measure the correction in the same way, finding $\text{fit}/\text{cont} = 0.857$. We rescale the \textit{PCANN-QSANNdRA} prediction by this correction in order to compare it to the true continuum.

The IGM absorption bias over the \lab forest at $2.0<z<3.0$ is far harder to estimate since no definitive comparable measurements in high-resolution spectra exist in the literature. The mid-point of \lab absorption redshift in our sample is $z=2.906$. We estimate a correction by assuming a fixed ratio $r_{\alpha\beta}$ of \lal and \lab optical depth in the IGM: $\tau_\beta \ (z=2.906) = r_{\alpha\beta} \ \tau_\alpha (z=2.906) + \tau_\alpha(z=2.297)$. In an optically-thin regime, $r_{\alpha\beta} = 0.173$ in each pixel (e.g.~\citealt{White03,Eilers19}). The ratio of effective (binned) optical depths is affected by the presence of temperature fluctuations and fluctuations in the UVB \citep{Oh05}. Assuming a power-law temperature-density relation $T\propto\Delta^{\sim 0.6}$ at $2.3<z<3.0$ \citep{Lee15}, we can estimate a ratio closer to $\tau_\beta \lesssim 0.333 \ \tau_\alpha$. Between these two extremes, the absorption correction is the range $\text{fit}/\text{cont} = 0.987 - 0.9291$. The measurements of \citet{Songaila04} are in closer agreement with the lower limit, implying a correction $\text{fit}/\text{cont} = 0.94 \pm 0.03$. We will measure the accuracy of the automatic continuum fitter over the \lab forest, and draw constraints on the properties of the IGM, using forward-modelled numerical simulations in future work. Until then, we adopt the mean of the bounds above and include overall rescaling uncertainties which span them.

\subsection{Bias and uncertainty estimations} \label{sec:bias}

Bias and uncertainty estimation is performed for all the continuum prediction methods listed in Section~\ref{sec:techniques} using the testing sample. More specifically, the \textit{Power-law}, \textit{PCA-P\^aris-10}, \textit{PCA-P\^aris-5} and \textit{PCA-Davies-smaller} are tested by fitting the model to the red side range (as defined slightly differently for each method) and predicting the blue side, then comparing with the continuum recovered by the automated blue-side fitter described in Section~\ref{sec:cont-fit}. In the \textit{Neighbours} method, each quasar in the testing sample draws its closest neighbours from the training sample. Finally, the \textit{PCANN-QSANNdRA} and \textit{PCA-Davies-nominal} methods are both trained on our training sample using their specified procedures.

The accuracy of machine learning techniques is often measured using cross-validation, during which the training procedure is repeated many times on resampled training and testing samples of the same size (e.g.~\citealt{Richards11-crossval}). Training the \textit{PCANN-QSANNdRA} and \textit{PCA-Davies-nominal} methods takes a prohibitively long time to perform cross-validation statistically. To estimate the effect of random selection of the samples, we train \textit{PCA-Davies-nominal} on the total sample (testing$+$training). We find the prediction scatter on the testing sample only decreases by $\lesssim0.1\%$ while the bias is unchanged. Since empirical prediction errors are much larger, we conclude the lack of cross-validation is not biasing our results.

The wavelength-dependent bias is computed as the mean deviation of the predictions from the automated blue-side fit at each wavelength. We record the asymmetric central $68$ and $95$ percentile intervals of the residual distribution as the $1\sigma$ and $2\sigma$ wavelength-dependent uncertainties.

\subsection{\lal and \lab opacities at $z>5.5$} \label{sec:highz}

We now apply our continuum reconstruction to a sample of $z > 5.5$ quasars. In order to cover $\lambda\leq2000$\AA \ in the rest-frame, quasars must have existing near infrared (NIR) spectra. We apply a depth criterion of SNR $>7$ on the spectra binned to eBOSS resolution to match our training samples. To limit the uncertainties arising from spectral reduction residuals which can differ between instruments, we restrict ourselves to using spectra taken with the X-Shooter spectrograph on the Very Large Telescope \citep{XSHOOTER} which has delivered the majority of NIR spectra of high-$z$ quasars. The selected quasars are listed in Table~\ref{tab:quasars}. All spectra were reduced using the open-source software \textit{PypeIt} \citep{pypeit, Pypeit-official} and were previously described in \citet{Meyer19} and \citet{Eilers20}.  Nine out of the $19$ spectra were also used in the measurements of \citet{Bosman18}, and a further $3$ quasars were used but without X-Shooter spectra. The $12$ quasars we are using here therefore made up $12/64$ sightlines in \citet{Bosman18}. There is also overlap with the sample of \citet{Eilers18}: $9/34$ of the quasars in that compilation are included here, albeit with IR spectra instead of the Echellette Spectrograph and Imager (Keck/ESI; \citealt{ESI}). The $7$ additional quasars from \citet{Eilers20} were pre-selected for having short proximity zones based on optical spectroscopy, which may indicate short lifetimes of the current quasar phase (see Appendix~\ref{app:prox}).

We masked known sky-lines, cosmic rays, and the region of high telluric absorption across $13450<\rev{\lambda_{\text{obs}}}(\text{\AA})<14250$ in observed wavelength from the following analysis. The spectra are processed differently depending on which continuum reconstruction method is used. For the \textit{Power-Law}, the parameters are fit directly to the un-binned X-Shooter spectrum without re-binning. To mimic the procedure of \citet{Paris11}, the X-Shooter spectra are re-binned to BOSS resolution and the flux uncertainties are added in quadrature. For the remaining techniques, the red sides of the X-Shooter spectra are fitted with a spline continuum with the same \rev{point spacing} in rest-frame velocity as our testing sample. The red-side continuum is fit with the PCA components of the \textit{PCA-P\^aris-5, PCA-Davies-nominal, PCA-Davies-smaller} and \textit{PCANN-QSANNdRA} model, while the $40$ closest neighbours of the quasar are drawn from the eBOSS training sample. 
\rev{Since all quasars in this work have broad emission lines which are well resolved both by X-Shooter and in eBOSS spectra, the different spectral resolutions of the instruments does not affect the spline-fitting procedure. 
}
The blue-side predictions are then divided by the mean bias curves shown in Figure~\ref{fig:compar}. The $1\sigma$ upper and lower bounds are obtained by using the upper and lower bounds containing $68\%$ of the observed scatter in the testing sample.

The mean transmitted flux is computed with respect to the intrinsic continuum predictions in fixed redshift intervals of $\Delta z = 0.2$. For each spectrum, we exclude pixels affected by sky-lines and transmission located within the proximity zone of the quasar, $\lambda>1176$\AA \ for \lal and $\lambda>1000$\AA \ for \lab \citep{Bosman18, Eilers18}. We show the effect of changing the limits of the proximity zone in the Appendix~\ref{app:prox}. Further, we use the automated red-side continuum fits to help in the manual identification of intervening metal absorbers. We fit the Doppler width $b$ of absorbers and exclude regions of the \lal and \lab forest within separations $\Delta v = 5b$ of metal absorbers, in an effort to mitigate the effect of DLAs in our sample \citep{Meyer19}. We note, however, that DLAs become increasingly metal-poor at $z>4$ \citep{Wolfe05, Rafelski14} and this process may be insufficient. $17\%$ of our $2.7<z<3.5$ training and testing samples had to be excluded due to the presence of DLAs. Table~\ref{tab:dz} shows the remaining effective $\int dz$ probed in each redshift bin by our sample and the effective redshift mid-points of non-excluded pixels.

\section{Results} \label{sec:results}

\subsection{Bias and Scatter}

The mean bias and $1,2\sigma$ contours of the scatter are shown in Figure~\ref{fig:compar}. We find the \textit{Power-Law} reconstructions are the most biased on average over the \lal forest, with $B_{\text{Ly}\alpha} = \text{Prediction}/\text{True}-1 = -9.58\% \ {-13.2/+13.1} \ {(-22.2/+32.0)}$. Looking at the wavelength-dependence of the bias, it becomes clear the offset is due to the power-law's inability to reproduce broad emission lines on the blue side. The continuum emission in-between the emission lines, however, is very accurately captured\rev{, except at the shortest wavelengths $\lambda<1025$\AA.}

Using all $10$ components from \citet{Paris11}, \textit{PCA-P\^aris-10} gives a much lower bias, $B_{\text{Ly}\alpha} = 2.66\% \ {-11.9/+11.7} \ {(-22.2/+28.9)}$. The low bias is in agreement with the flux calibration tests conducted in \citet{Paris11}, but we find a scatter about twice larger, likely due to the small size of the original sample. We find that restricting the PCA to $5$ components results in worse performance, with the largest scatter of any method at $B_{\text{Ly}\alpha} = 8.22\% \ {-13.6/+13.3} \ {(-25.2/+32.1)}$. One possible reason is that the lack of re-training of the smaller PCA, since the projection matrix between red-side and blue-side components is simply truncated. Components beyond the $5^{\text{th}}$ likely encode, on average, a mean correction to the continuum. The older PCAs and the power-law reconstruction are both outdone by stacks of \textit{Neighbours}, which have accuracies $B_{\text{Ly}\alpha} = 1.9\% \ {-10.3/+10.2} \ {(-19.1/+24.6)}$.

Both newer PCA techniques, trained on larger samples, perform better than stacking \textit{Neighbours}. \textit{PCANN-QSANNdRA} displays scatter below $10\%$ at $1\sigma$, with $B_{\text{Ly}\alpha} = 0.3\% \ {-9.2/+9.1} \ {(-18.1/+22.9)}$. Despite employing a very different architecture, \textit{PCA-Davies-nominal} performs very similarly with scatter just below $9\%$: $B_{\text{Ly}\alpha} = 0.9\% \ {-9.0/+8.8} \ {(-17.1/+21.4)}$. The increased sample size of the training sample, better purity, and improved intrinsic continuum recovery compared to \textit{PCA-Davies-smaller} makes a clear difference, as the latter has $B_{\text{Ly}\alpha} = 4.7\% \ {-10.4/+10.1} \ {(-19.5/+24.8)}$.

Finally, using only wavelengths which would be observed in the optical at $5.5<z<6.2$, \textit{PCA-optical only} predictably performs worse than new PCAs and stacks of neighbours with access to longer wavelengths, but still better than both older PCAs and power-law reconstructions, with $B_{\text{Ly}\alpha} = 1.0\% \ {-11.3/+11.2} \ {(-20.2/+27.5)}$.

Over the \lab forest ($972<\lambda(\text{\AA})<1000$) the distinction in performance between methods is similar, with the \textit{Power-Law} method doing the worst with $B_{\text{Ly}\beta} = -12.5\% \ {-20.1/+19.9} \ {(-31.7/+50.8)}$ at $1\sigma \ (2\sigma)$. Again, both larger-sample PCAs are more accurate with $B_{\text{Ly}\beta} = 0.4\% \ {-17.2/+16.3} \ {(-31.1/+40.5)}$ for \textit{PCA-Davies-smaller}. The improved training sample improved the bias and reduced scatter below $17\%$ to $B_{\text{Ly}\beta} = 1.4\% \ {-16.5/+15.5} \ {(-30.6/+39.1)}$ for \textit{PCA-Davies-nominal}, over-taking \textit{Neighbours} stacking which has accuracy $B_{\text{Ly}\beta} = 3.4\% \ {-17.6/+17.0} \ {(-31.5/+42.0)}$. \textit{PCANN-QSANNdRA} very similarly to \textit{PCA-Davies-nominal}, with accuracy $B_{\text{Ly}\beta} = 3.8\% \ {-16.8/+16.1} \ {(-31.3/+39.9)}$. 
Finally, the optical wavelengths-only PCA performs significantly worse than IR PCAs but still better than power-law extrapolation, with $B_{\text{Ly}\beta} = 1.7\% \ {-18.3/+17.7} \ {(-32.2/+45.8)}$.

Beyond mean biases, the residuals from the different methods show very different wavelength dependences (Figure~\ref{fig:compar}). Residuals at the location of broad emission lines can be seen in the \textit{Power-Law}, \textit{PCA-P\^aris-10} and \textit{PCA-P\^aris-5} methods, while the \textit{Neighbours} technique and PCA methods trained on larger samples (\textit{PCA-Davies-smaller}, \textit{PCA-Davies-nominal}) do not show residuals at the locations of the N~{\small{II}} 1085\AA, Fe~{\small{III}} 1125\AA \ and C~{\small{III}} 1175\AA \ broad emission lines. This indicates that correlations between blue-side emission lines and red-side properties are, unsurprisingly, adequately captured by stacks of quasars with similar red-side properties. The more recent PCA decompositions also capture broad-line correlations.

An important addition in the \textit{PCA-Davies-smaller} and \textit{PCA-Davies-nominal} was an independent fitting parameter $|\delta z|<0.05$ to account for redshift uncertainties in the testing sample. Indeed, small residuals visible around the O~{\small{VI}} line in \textit{PCANN-QSANNdRA} are likely due to the lack of redshift correction as a fitting parameter, which is mitigated by the larger number of components ($53/42$ for the red/blue sides compared to $10/6$). The absence of a redshift shift parameter in the other models might disadvantage them if the quasar redshifts in eBOSS are more inaccurate than in $z>5.5$ quasars. Indeed, the systemic redshifts of high-$z$ quasars are often known to within $|\delta z|<0.0005$ from detections of sub-millimeter emission lines in quasar hosts \citep{Willott15, Decarli17,Eilers20}. The $n>5$ components of the \textit{P\^{a}ris-10} PCA introduce shape perturbations at the edges of broad emission lines which account for some of these redshift uncertainties. We find that these high-order components give continua unphysical shapes (see e.g.~\citealt{Kakiichi18}) but they could potentially perform better in the absence of redshift uncertainties.

All wavelength-dependent biases show an upturn of uncertainties in the \lab region, which could be intrinsic or artificial. Turn-overs in the NUV SEDs of quasars have been observed around $\lambda\sim1000$\AA \ in $z<2$ quasars not significantly affected by IGM absorption (see Section~\ref{sec:PL}). If the FUV slopes of quasars are determined by processes distinct from the NUV, the deviation may be intrinsic. Indeed, even stacks of nearest neighbours which match NUV properties very closely display increased errors at $\lambda<1000$\AA \ (Figure~\ref{fig:compar}). 
Alternatively, increased uncertainties could originate from the performance of our intrinsic-continuum recovery algorithm being strongly redshift-dependent. We have estimated the mean bias due to IGM absorption in the \lab forest, but our sample is too small and the $z>3$ true \lab absorption too poorly constrained to quantify the redshift-dependent bias. It is possible that our algorithm performs significantly worse on $z>3$ quasars with more absorption, which would drive up scatter artificially. 
Unfortunately, this is a limitation of currently available $z<3.5$ testing samples and measurement of the intermediate-redshift \lab forest. Future large spectroscopic samples of $z<2$ quasars with UV coverage, or upcoming surveys of quasar spectra at $z>3$ at higher resolution and SNR than eBOSS such as the WHT Enhanced Area Velocity Explorer (WEAVE, \citealt{WEAVE}) and the Dark Energy Spectroscopic Instrument survey (DESI, \citealt{DESI}) will help resolve this issue in the future.

\section{Updated \lal and \lab opacities at $z>5.5$}\label{sec:highz-res}

\begin{figure}
\includegraphics[width=0.5\textwidth]{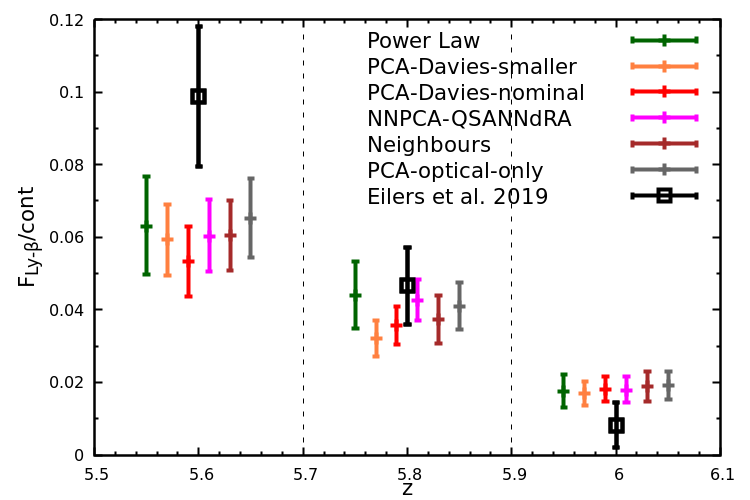}
\caption{Mean transmitted \lab flux along the sightlines of quasars at $z>5.7$. The coloured points correspond to predictions made from the five reconstruction methods which extend to $\lambda=972$\AA, with bias corrections applied from testing on the $2.7<z<3.5$ eBOSS testing sample. The uncertainties account for both the $1\sigma$ uncertainties in reconstruction accuracy and observational errors. All techniques agree within $1.5\sigma$.}
\label{fig:lyb}
\end{figure}

The mean \lal and \lab fluxes recovered at $4.7<z<6.3$ after applying wavelength-dependent bias corrections and accounting for reconstruction uncertainties are given in Table~\ref{tab:dz} for the $2$ best continuum reconstruction methods, \textit{PCA-Davies-nominal} and \textit{PCANN-QSANNdRA}. The techniques agree within $1.5\sigma$, and also agree with the remaining reconstruction methods within $1.5\sigma$ at all redshifts. Results are shown in Figure~\ref{fig:lya} and Figure~\ref{fig:lyb}.

We compute errors on the mean using continuum reconstruction uncertainties and observational flux uncertainties only, not accounting for cosmic variance. Mean transmitted fluxes are crucial for quantitatively understanding the end stages of reionisation and calibrating numerical simulations (e.g.~\citealt{Kulkarni19, Keating20}). Previous measurements of $z>5.0$ transmission binned the data in $50$ cMpc chunks before computing the mean and scatter based on the centres of each bin, leading to flux outside the stated redshift ranges being implicitly included. Our definition removes the dependence on the distribution of quasar redshifts in the sample introduced by this procedure, as we simply weigh sightlines by the length of usable transmission (see also \citealt{Worseck19}).

The mean \lal transmitted fluxes differ from previous studies significantly. At $z=5.2$ ($z=5.4, 5.6$), we find values differing at $-1.5\sigma$ ($-1.3\sigma, -2.5\sigma$) from the study of \citet{Bosman18} which used purely power-law reconstructions. We re-bin our fluxes following the definition of \citet{Eilers18} and find similar but opposite discrepancies of $+1.8\sigma$ ($+2.2\sigma, +2.6\sigma$) at $z=5.25$ ($z=5.5, z=5.75$). The origin of the discrepancy is therefore likely to be the opposite systematic biases in the \textit{Power-Law} reconstruction and the multi-PCA method used in \citet{Eilers18}, which we approximated with the \textit{PCA-P\^aris-5} technique.  After applying the corresponding systematic biases calculated in the previous Section to the transmitted flux values in the literature, the measurements agree with our new values within $1.5\sigma$ in all bins except the $z=5.75$ bin of \citet{Eilers18}. However, one must also be mindful of the large cosmic variance between sightlines at these redshifts; only $9$ of the $19$ sightlines we used were included in both previous studies.

Because cosmic variance is known to be large at $z\sim5.5$, we make no attempt to quantify it using our sample of only $19$ sightlines. Indeed, the extremely rare opaque sightline of the quasar J0148+0600 \citep{Becker15} accounts for $1/19^\text{th}$ of our sample, while it made up only $1/64^\text{th}$ of the \citet{Bosman18} sample. It is also very difficult to estimate how many sightlines would be necessary to constrain cosmic variance. 
Based solely on continuum uncertainties, $19$ quasar sightlines can constrain the mean fluxes below $4\%$ ($8\%$) using power-law reconstructions at $1\sigma$ ($2\sigma$). By using the \textit{PCA-Davies-nominal} method, this can be improved to $\sim2\%$ ($5\%$). 
%Based on the distribution of intrinsic transmission reported in \citet{Bosman18} and \citet{Eilers18} and the occurrence rates of extremely opaque sightlines, we estimate the additional uncertainty to our smaller sample size due to cosmic variance to be of the order of $\pm 2-5 \%$. Assuming the same quasar spectra SNR distribution, we estimate it would require $\sim50$ quasar sightlines to get constraints below $5\%$ after accounting for cosmic variance. 

%While samples of over $50$ sightlines have been analysed in similar ways, the majority of quasar spectra were only observed in optical wavelengths, which introduces a number of caveats. The PCA decomposition of \citet{Paris11} was computed out to $2000$\AA \ in the rest frame and is not straightforwardly applicable to optical-only high-$z$ spectra. The samples of \citet{Bosman18} and \citet{Eilers18} contain $62$ and $34$ sightlines, the vast majority of which were observed in the optical. Even after applying our \textit{Power-Law} and \textit{PCA-P\^aris-5} systematic bias corrections, which were the least accurate out of the models we tested, the transmitted flux values at $z=5.2$ and $z=5.25$ in the former and latter still disagree at $>2\sigma$ (but note there is also a difference in definition due to binning as described previously).
We note that since the power law's bias is dominated by broad line residuals, its predictions are not immune to intrinsic quasar evolution with redshift. Indeed, if $z>5.7$ quasars have systematically weaker broad lines on both the red and blue side, the bias correction for the power-law fitting would be systematically over-estimated (see top left panel of Figure~\ref{fig:compar}) or in other words, power-law extrapolations would be more correct on average at higher redshift. Perhaps surprisingly, we find that a PCA trained using only the $1220<\lambda(\text{\AA})<1450$ wavelength range succeeds in removing wavelength-dependent residuals coming from broad emission lines on the blue side (Figure~\ref{fig:compar}, bottom right panel). The optical-only PCA has scatter $11.3\%$ and a small bias of $1.0\%$ over the testing sample. Theoretically, this implies the mean \lal transmission could be constrained without bias with a sufficient number of quasar spectra with appropriate SNR covering only observed optical wavelengths.

%However, we note that the values of \lal transmitted flux recovered from our $z>5.7$ sample using the \textit{Power-Law} and \textit{PCA-optical only} techniques are consistently larger than using all other techniques by $\sim1.5\sigma$ after accounting for reconstruction bias and scatter. This seems to suggest that the high-$z$ sample differs from the low-$z$ sample at fixed $1220<\lambda(\text{AA})<1450$ properties. One possible reason could be evolution of the broad emission lines C~{\small{IV}} and C~{\small{III}}], which are those closest in ionisation potential to emission lines on the blue side. The residuals of the optical-only PCA, however, do not show any features related to broad line on the blue side, implying their properties are correctly recovered in the testing sample. Alternatively, the tension could be caused by the testing sample and high-$z$ quasars having different luminosity distributions, since this is known to correlate with the quasar NUV SED slope. Quantifying the effect of luminosity on reconstruction would require using low-$z$ quasars with matched luminosity, which is currently impractical with eBOSS.

Over \lab transmission, all methods have scatters of at least $16\%$ and systematic biases as large as $12\%$. Despite this, the recovered transmitted fluxes agree within $1.5\sigma$ for all methods after wavelength-dependent bias corrections were applied, and reconstruction uncertainties are taken into account (Figure~\ref{fig:lyb}). We find a shallower evolution of \lab transmitted flux than reported in \citet{Eilers19} with a discrepancy of \rev{$\sim1.5\sigma$} at $5.5<z<5.7$. \rev{The \lab continuum reconstruction method used in \citet{Eilers19} stitches a PCA akin to \textit{PCA-P\^aris-5} onto a quasar continuum from \citet{Shull12}, which is a complex procedure giving rise to unknown but likely significant bias and uncertainty. However, cosmic variance could also be the main driver of the observed discrepancies since the quasar samples of the two studies do not overlap.
}

\subsection{Caveats}

Three final caveats, which we do not address in this paper, relate to spectroscopic reduction residuals and selective wavelength masking. We outline those issues here and reflect on their consequences and possible mitigation measures. 

The X-Shooter spectrograph has a data reduction process far more complex than eBOSS spectra, which involves an order of magnitude more stitchings of spectral orders in the observed IR and combination of the optical and IR spectrograph arms. Such residuals can bias all reconstruction methods, and perhaps impacting most strongly PCA methods as they aim to fit detailed red side features and least strongly neighbour-stacking methods which will implicitly marginalise over defects. Quantifying the effect of reduction errors could potentially be done by using large samples of quasars observed with both X-Shooter and the BOSS spectrograph, such as the XQ-100 sample \citep{Lopez16}.

Second, the rest-frame wavelengths affected by telluric absorption depend directly on the redshift of observed quasars. In this work, we masked the regions most severely affected by telluric absorption. All methods can implicitly adapt to missing data, but it may have a deeper effect on PCA methods. Since the number of PCA components is chosen to contain a fixed fraction of the observed variance in the training sample over a complete wavelength range, the same components might not have been retained if the same wavelength regions are missing. To quantify this effect, one could construct `tailored' PCA decompositions by extracting the component vectors from a training sample in which the same wavelength regions are always masked.

Finally, the potential impact of quasar evolution with redshift on PCA reconstruction will be addressed in future work. Quasars with fewer neighbours will instinctively be less well captured by PCA reconstructions, and may occupy sparsely-populated regions of PCA component coefficients parameter space. Different definitions of quasar distance, and `tailored' PCAs, may be ways to explore these effect in the future.

\section{Conclusions}\label{sec:ccl}

We have tested eight different quasar continuum reconstruction methods designed to measure \lal and \lab transmission at $z>5.0$. We use a common testing sample consisting of $4579$ quasars at $2.7<z<3.5$ from the eBOSS catalogue. For three techniques which require training at low-$z$, we use a separate common training sample consisting of the same number of quasars. Our findings are:\begin{itemize}
	\item The continuum uncertainties arising from using power-law extrapolation are larger than has previously been assumed, at $+13.2\%/-13.1\%$ over \lal and  $+19.9\%/-20.1\%$ over \lab.
	\item Power-law reconstructions are also the most systematically biased due to the presence of broad emission lines on the blue side, by $\geq-9.58\%$ over \lal.
	\item PCA reconstructions trained on small samples and restricted to fewer components are also significantly biased, by $\geq+8.22\%$ over \lal.
	\item Using PCAs trained on large, cleaner samples and using improved blue-side recovery techniques, which we present there, it is possible to reduce the reconstruction uncertainty below $9\%$ over \lal and below $17\%$ over \lab. 
	\item The neural-network-mapped PCA, \textit{PCANN-QSANNdRA}, performs nearly identically to newer `classical' PCA methods for the $1\sigma$ scatter over the \lal and \lab forests.
	\item Power-law reconstructions and small-sample PCAs create strong wavelength-dependent biases and scatter, of which newer methods are devoid.
\end{itemize}

We then assembled a sample of $19$ X-Shooter spectra of $z>5.7$ quasars with eBOSS-equivalent SNR $>7$ and present new results for the \lal and \lab transmission at $z>5.0$, correcting for the wavelength-dependent biases we identified and accounting for continuum uncertainties. 
We conclude that:\begin{itemize}
	\item In the absence of rigorous tests at $z<5$, measurements of \lal and \lab transmission at $z>5.0$ are liable to be biased by inaccurate continuum reconstructions by systematic errors in excess of statistical uncertainties.
	\item Using $z>5.7$ quasar spectra covering only observed optical wavelengths, it is still possible to constrain the intrinsic continuum within $+11.2\%/-11.3\%$ using optical-only PCA. 
	\item Since the bias of power-law reconstructions is driven by broad emission lines on the blue side, we caution that they are not necessarily exempt from caveats related to intrinsic quasar evolution with redshift, whether due to a mismatch in intrinsic luminosity or more complex factors.
\end{itemize}

As more complex techniques will keep improving the quality of continuum predictions, it is important to keep testing these methods on well-controlled samples at low-$z$. The number of known quasars at $z>5.7$ and the quality of their spectra is quickly increasing, and reconstruction techniques for quasar continuum emission are likely to remain a crucial tool in the study of reionisation. 

\section*{Acknowledgements}

The authors thank George Becker and Romain Meyer for providing useful and insightful comments on the manuscript. SEIB  acknowledges  funding  from  the  European  Research Council  (ERC)  under  the  European  Union's  Horizon2020 research and innovation programme (grant agreements No.~669253 `First Light' \rev{and No.~740246 `Cosmic Gas'}). ACE acknowledges support by NASA through the NASA Hubble Fellowship grant\#HF2-51434  awarded  by  the  Space  Telescope  Science  Institute,  which  is  operated by the Association of Universities for Research in Astronomy, Inc., for NASA, under contract NAS5-26555.

This research has made use of NASA's Astrophysics Data System, and open-source projects including \texttt{ipython} \citep{ipython}, \texttt{scipy} \citep{scipy}, \texttt{numpy} \citep{numpy}, \texttt{astropy} \citep{astropy1,astropy2}, \texttt{scikit-learn} \citep{scikit-learn} and \texttt{matplotlib} \citep{matplotlib}.

Funding for the Sloan Digital Sky Survey IV has been provided by the Alfred P.~Sloan Foundation, the U.S.~Department of Energy Office of Science, and the Participating Institutions. SDSS acknowledges support and resources from the Center for High-Performance Computing at the University of Utah. The SDSS web site is \url{www.sdss.org}.

SDSS is managed by the Astrophysical Research Consortium for the Participating Institutions of the SDSS Collaboration including the Brazilian Participation Group, the Carnegie Institution for Science, Carnegie Mellon University, the Chilean Participation Group, the French Participation Group, Harvard-Smithsonian Center for Astrophysics, Instituto de Astrof\'isica de Canarias, The Johns Hopkins University, Kavli Institute for the Physics and Mathematics of the Universe (IPMU) / University of Tokyo, the Korean Participation Group, Lawrence Berkeley National Laboratory, Leibniz Institut f\"ur Astrophysik Potsdam (AIP), Max-Planck-Institut f\"ur Astronomie (MPIA Heidelberg), Max-Planck-Institut f\"ur Astrophysik (MPA Garching), Max-Planck-Institut f\"ur Extraterrestrische Physik (MPE), National Astronomical Observatories of China, New Mexico State University, New York University, University of Notre Dame, Observat\'orio Nacional / MCTI, The Ohio State University, Pennsylvania State University, Shanghai Astronomical Observatory, United Kingdom Participation Group, Universidad Nacional Aut\'onoma de M\'exico, University of Arizona, University of Colorado Boulder, University of Oxford, University of Portsmouth, University of Utah, University of Virginia, University of Washington, University of Wisconsin, Vanderbilt University, and Yale University.

\section*{Data availability}

The data underlying the testing and training samples, and further examples of the fitting methods, are available on the lead author's website at \url{http://www.sarahbosman.co.uk/research}. The remainder of the data underlying this paper will be shared upon request to the corresponding author.

%%%%%%%%%%%%%%%%%%%%%%%%%%%%%%%%%%%%%%%%%%%%%%%%%%



\bibliographystyle{mnras}
\bibliography{bibliography} % if your bibtex file is called example.bib

%\begin{thebibliography}{99}
%\end{thebibliography}

%%%%%%%%%%%%%%%%%%%%%%%%%%%%%%%%%%%%%%%%%%%%%%%%%%

%%%%%%%%%%%%%%%%% APPENDICES %%%%%%%%%%%%%%%%%%%%%

\appendix

\section{Dependence of Power-Law reconstruction on wavelength range}\label{app:PL}

\begin{figure}
\includegraphics[width=0.5\textwidth]{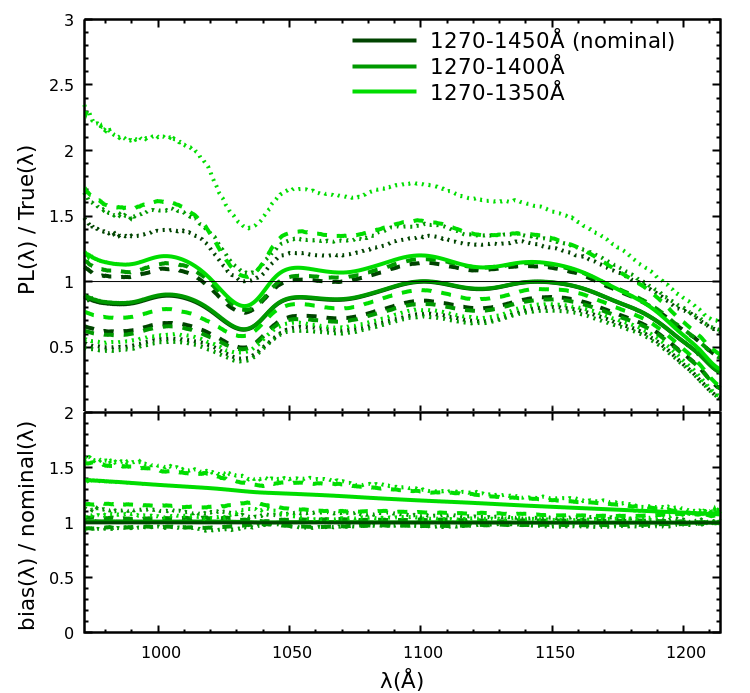}
\caption{The effect of varying the wavelength range over which a power-law reconstruction is fit before extrapolating to the blue side. Dashed and dotted lines correspond to the contours encompassing $86\%$ (1$\sigma$) and $95\%$ ($2\sigma$) of the scatter, respectively.}
\label{fig:app-pl}
\end{figure}

Depending on quasar redshift, restricted ranges in wavelengths have to be used to fit power-law reconstructions at $z>5.7$ when only optical observations are available. We test the effect of varying the fitting range by repeating the procedure in Section~\ref{sec:PL} and changing that parameter only. The results are shown in Figure~\ref{fig:app-pl}. Using a shorter range $1270<\lambda(\text{\AA})<1400$ does not produce any significant extra bias compared to the nominal range $1270<\lambda(\text{\AA})<1450$ we have used throughout the paper. The prediction scatter is increased by the shorter range by $\pm2\%$ ($\pm4\%$) at $1\sigma$ ($2\sigma$).

Shortening the range further to $1270<\lambda(\text{\AA})<1350$ does introduce an extra bias of $+16\%$ and dramatically inflates the scatter ($+49\% / -36\%$ at $1\sigma$). We suspect this is because the presence of the weak Si~{\small{II}} $1303$\AA \ and broad N~{\small{V}} emission lines have a stronger effect on the best fit in the absence of the up-turn towards the broad Si~{\small{IV}} $1396$ \AA \ line. With a smaller fitting range, it is harder to separate the broad lines from the continuum and the fit becomes influenced by line ratios, resulting in increased bias and much increased scatter.

\section{Effect of Proximity Zone Cut-Off on recovered $z>5$ transmission}\label{app:prox}

A potentially important free parameter is the wavelength which marks the end of quasars' effect on the surrounding IGM, $\lambda_\text{prox}$. In \citet{Bosman18}, we showed that stacks of quasar spectra at $z>5.7$ display no boost beyond $\lambda_\text{prox} = 1176$\AA \ and therefore adopted the value. We now wish to re-visit this measurement with accurate continuum reconstruction rather than the directly measured flux. According to the residuals in the power-law in Figure~\ref{fig:compar}, the bias of the power-law fit at $\lambda>1176$\AA \ at $z<3.5$ is not due only to a quasar's proximity zone, but also to very broad and/or blue-shifted \lal emission lines. Indeed, the \textit{PCA-Davies} series of PCAs show no bias until $\lambda \leq 1190$\AA. Either way, it may be possible to recover IGM transmission to larger wavelengths than previously assumed. This is crucial, since the $5.9<z<6.1$ interval is only probed by a total $\int dz = 3.888$ using $\lambda_\text{prox} = 1176$\AA. Furthermore, the stack of $z>6.1$ quasars in \citet{Bosman18} shows much shorter proximity zones on average, only affecting the transmission at $\lambda\gtrsim1190$\AA.

\begin{figure}
\includegraphics[width=0.5\textwidth]{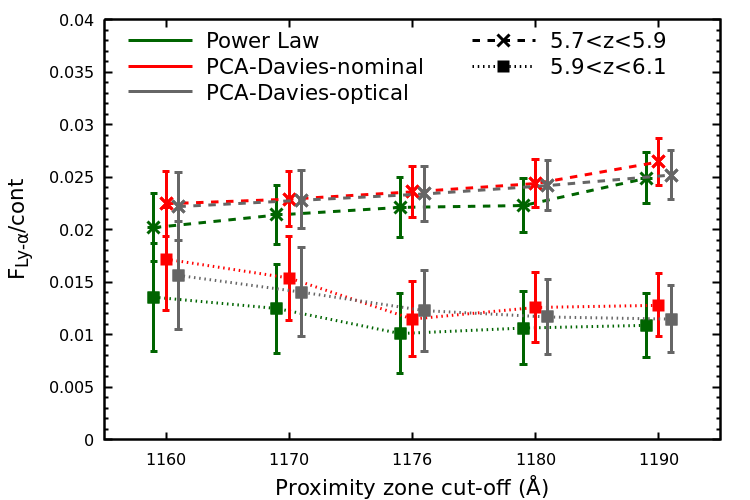}
\caption{Values of mean transmitted \lal flux at $5.7<z<5.9$ and $5.9<z<6.1$, showing the effect of varying the proximity zone cut-off wavelength. We only show three methods, \textit{Power-Law}, \textit{PCA-Davies-nominal} and \textit{PCA-optical only} for legibility. The transmitted flux increases as the cut-off is shifted to $\lambda_\text{prox} = 1190$\AA \ at $5.7<z<5.9$, reflecting the mean enhancement in quasars' proximity zones. The effect is not seen at $5.9<z<6.1$, where proximity zones are shorter. The increase towards $\lambda_\text{prox} = 1160$\AA\ at $5.9<z<6.1$ is due to shifting of the mean redshift and reduced sample size.}
\label{fig:app-prox}
\end{figure}

We therefore repeat the procedure in Section~\ref{sec:highz-res} varying only $\lambda_\text{prox}$. The results are shown in Figure~\ref{fig:app-prox} for two redshift ranges, $5.7<z<5.9$ and $5.9<z<6.1$.  We find an increase in transmission from $\lambda_\text{prox}=1176$\AA \ to $\lambda_\text{prox}=1190$\AA \ at $5.7<z<5.9$, but not $5.9<z<6.1$. This is in agreement with \citet{Bosman18}, and reflects the shorter sizes of proximity zones in $z>6.1$ quasars. In contrast, the transmission at $5.9<z<6.1$ shows an increase as the cut-off is made \textit{more} stringent, $\lambda_\text{prox}=1160$\AA. More stringent cut-offs severely reduce the total $\int dz$ probed by the sample from $\int dz = 6.55$ with $\lambda_\text{prox}=1190$\AA \ to $\int dz = 2.40$ with $\lambda_\text{prox}=1160$\AA, resulting in a large increase in uncertainty. Further, a stringent cut leads to a drift in the mean redshift of the transmission contributing to the bin, from $z_\text{mean} = 5.986$ with $\lambda_\text{prox}=1190$\AA \ to $z_\text{mean} = 5.975$ with $\lambda_\text{prox}=1160$\AA. Because the evolution with redshift is very fast, this leads to artificially inflated transmission.

We conclude the choice of $\lambda_\text{prox}$ can bias the measurement of mean transmission if it too loose but also if it is too stringent, when the sample size of quasars is small. Our choice of $\lambda=1176$\AA \ sits at an acceptable compromise of the two effects described above. $\lambda=1180$\AA \ might be a better choice as it increases the pathlength to $\int z = 4.578$ at $5.9<z<6.1$, an increase of $18\%$. However, we stick with $\lambda=1176$\AA \ due to the small increase in transmission towards $\lambda=1180$\AA \ at $5.7<z<5.9$ and the existence of individual $z>5.7$ quasars with proximity zones existing almost to $\lambda=1180$\AA \ \citep{Carilli10, Eilers19, Bosman20}.

\section{Improved Continuum Fitting}\label{app:contfit}

Figure~\ref{fig:app-contcompar} shows our improved spline-fitting procedure over the \lal and \lab forest. By explicitly removing strong absorbers, the fit is better able to recover the continuum with less bias over strong absorbers. This makes a difference at $z>3$, where our measured mean \lal flux is less biased than measured by \citet{Dallaglio09}, who used the original fitting procedure without the extra masking step.

\begin{figure*}
\includegraphics[width=\textwidth]{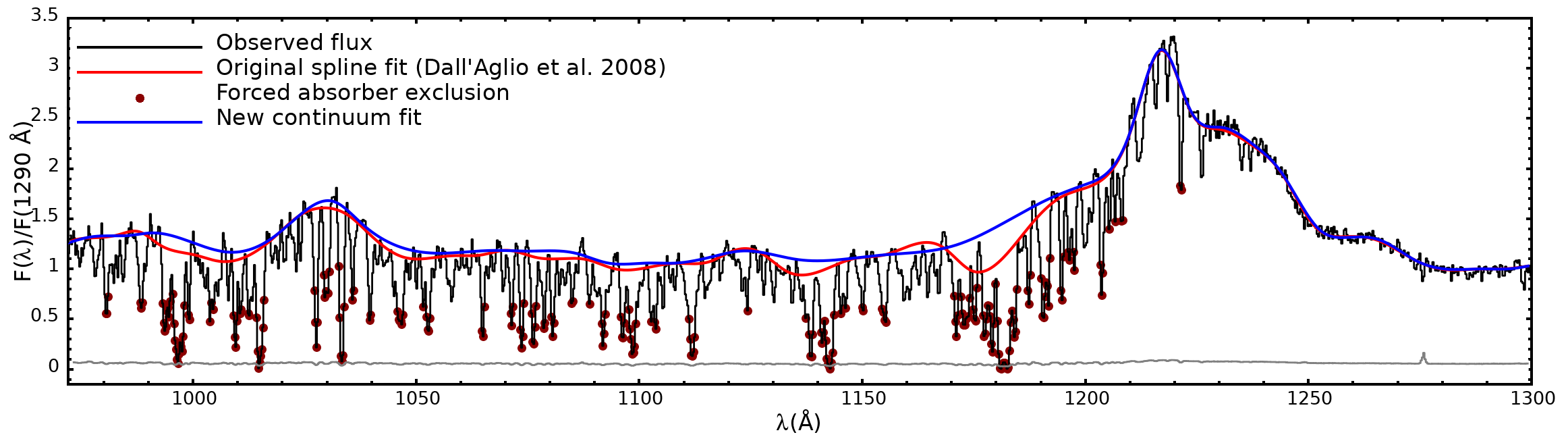}\label{fig:app-contcompar}
\caption{Showcase of the improved continuum fitting with exclusion of strong absorbers on the eBOSS quasar J113102.18+225910.5 at $z=3.373$, which is part of the testing sample.}
\label{fig:app-contcompar}
\end{figure*}

\section{\rev{Effect of SNR on reconstructions}}\label{app:snr}

\rev{We investigate the effect of SNR on the continuum reconstruction by dividing the testing sample into the lowest-SNR half (SNR$<10.5$) and highest-SNR half (SNR$>10.5$). We focus on the two best-performing methods: \textit{PCA-Davies-nominal} and \textit{PCANN-QSANNdRA}. For \textit{PCA-Davies-nominal}, the $1\sigma$ scatter increases slightly for the SNR$<10.5$ objects while the bias is slightly smaller. The nominal accuracy is $B_{\text{Ly}\alpha} = 0.9\% -9.0/+8.8$, the high-SNR sample yields $B_{\text{Ly}\alpha} = 1.0\% -8.4/+8.2$, and the low-SNR sample yields $B_{\text{Ly}\alpha} = 0.8\% -9.5/+9.3$. The same trend is observed with \textit{PCANN-QSANNdRA}: the $1\sigma$ uncertainties are increased by about $5\%$ for the lowest-SNR objects. In this case, the nominal accuracy is $B_{\text{Ly}\alpha} = 0.3\% -9.2/+9.1$, the high-SNR sample yields $B_{\text{Ly}\alpha} = 0.4\% -8.7/+8.6$, and the low-SNR sample yields $B_{\text{Ly}\alpha} = 0.2\% -9.7/+9.6$. }

\begin{figure*}
\includegraphics[width=\textwidth]{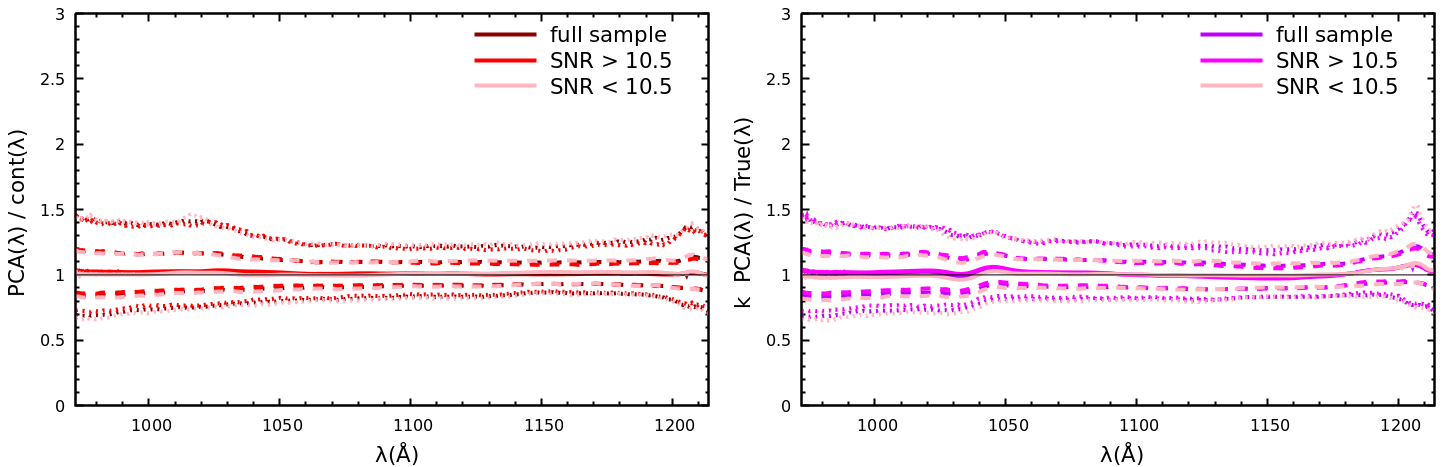}
\caption{\rev{Wavelength-dependent bias and scatter as a function of object SNR, for \textit{PCA-Davies-nominal} (left) and \textit{PCANN-QSANNdRA} (right). Dashed and dotted lines correspond to the contours encompassing $86\%$ (1$\sigma$) and $95\%$ ($2\sigma$) of the scatter, respectively. Dividing the testing sample into the highest-SNR half (SNR$>10.5$) and lower-SNR half (SNR$<10.5$) does not reveal significant effects in the mean uncertainty and bias, nor in their wavelength-dependence.}}
\label{fig:app-snr}
\end{figure*}

\rev{The slightly increased uncertainties on low-SNR objects could be due to larger continuum fitting uncertainties on both the red and blue sides. This effect does not affect the relative performance of the methods, as \textit{PCA-Davies-nominal} slightly outperforms \textit{PCANN-QSANNdRA} on both the high- and low-SNR samples. Lower SNR does not introduce extra bias in the predictions, neither in the mean bias nor in the wavelength-dependent bias (Figure~\ref{fig:app-snr}).}

\section{All fits}\label{app:allfits}

%This Section will be online-only.
We show below all bias-corrected continuum reconstructions for our $19$ $z>5.7$ quasars (Figure E1). We always mask the region of strong telluric absorption at $13450<\lambda(\text{\AA})<14250$ since telluric correction introduces residuals which are not present in our testing sample. In low-SNR ($\lesssim 1$/pixel) X-Shooter spectra, it is difficult to correctly model the response curve of the last order in the optical arm and the first order in the IR arm. When stitching the arms introduced artifacts, or when unfortunately-located strong absorbers coincided with the overlap, we opt to mask the affected region entirely. In P239-07 and P359-06, we also mask strong associated absorbers near the \lal emission line.

We do not make predictions for the quasars' \lal and \lab proximity zones, shaded in grey. The correction due to mean IGM absorption at $z<3.5$ is different over the \lal and \lab forest regions, leading to a discontinuity in the \lab proximity zone. Most fits are satisfactory, with the methods showing scatter with respect to each other but no clear bias. Below are some notes on individual objects.

\subsection{J0100+2802 and J1319+0950}
The wavelength-dependent bias correction applied to the \textit{Power-Law} fit reflect the average strength of blue-side broad emission lines at $z<3.5$. In J0100+2802 and J1319+0950, this mean prediction differs significantly from all other methods, indicating the quasars very likely have weaker O~{\small{VI}} emission lines than average. Both also display weaker than average lines on the red side.

\subsection{P056-16}
Interestingly, in P056-16, all methods make very similar predictions except for the  \textit{Power-Law}. All techniques predict a break in the power law continuum, including stacks of \textit{Neighbours} and {PCA-optical only}.

\subsection{J0818+1722}
The red side continuum of J0818+1722 is not well fit by any model. This is perhaps concerning since its stranger features (exceedingly broad and shifted C~{\small{IV}} emission line, extended C~{\small{III}}] and Si~{\small{IV}} lines) are similar to what is seen in the $z=7.5$ quasar \citep{Banados18}. Modifying the redshift by up to $\delta z = 0.1$, or using more complex convergence techniques to fit the PCA on the red side, did not alleviate this problem significantly. We plan to address this issue, and especially the dearth of low-$z$ analogues for certain $z>5.7$ quasars, in future work.

\begin{figure*}
\caption{All continuum predictions for the $z>5.7$ quasars used to measure the mean \lal and \lab opacity (including bias corrections). Reconstruction methods are indicated by different colours as shown in the legend. Notes on individual objects can be found in Appendix E.}
\includegraphics[width=\textwidth]{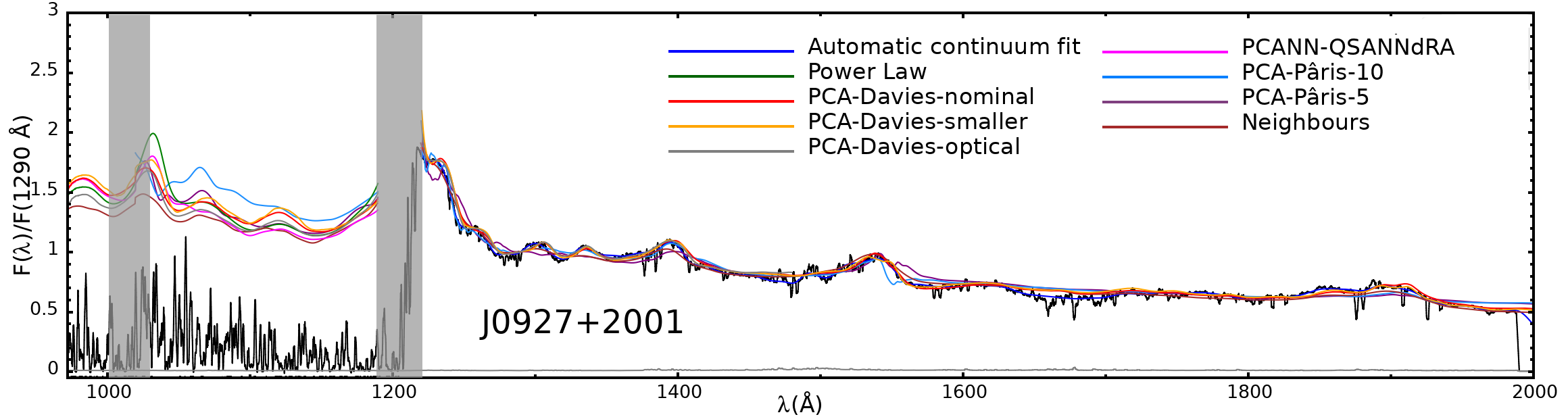}
\includegraphics[width=\textwidth]{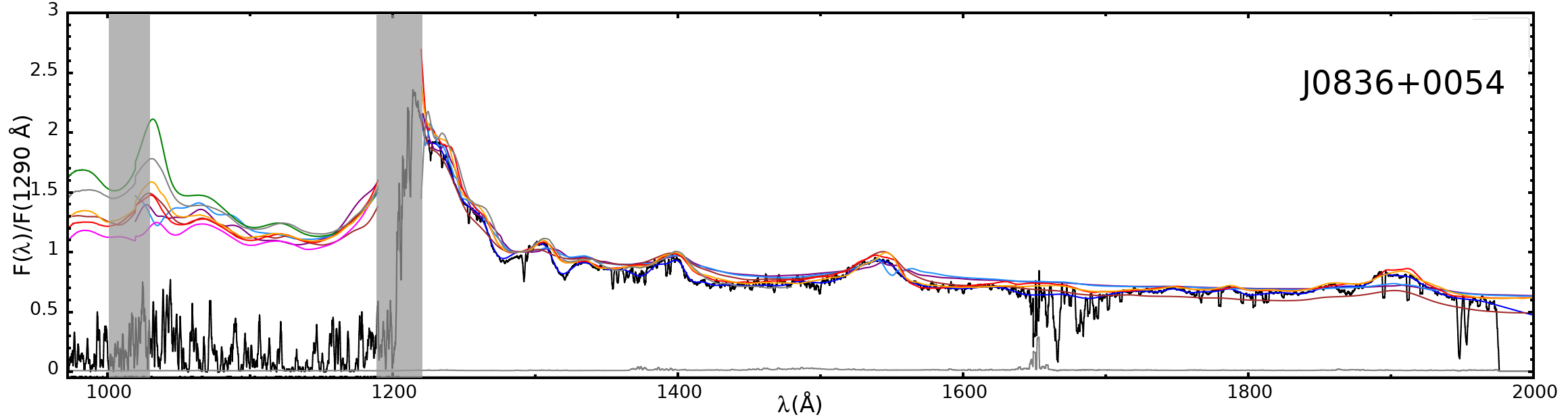}
\includegraphics[width=\textwidth]{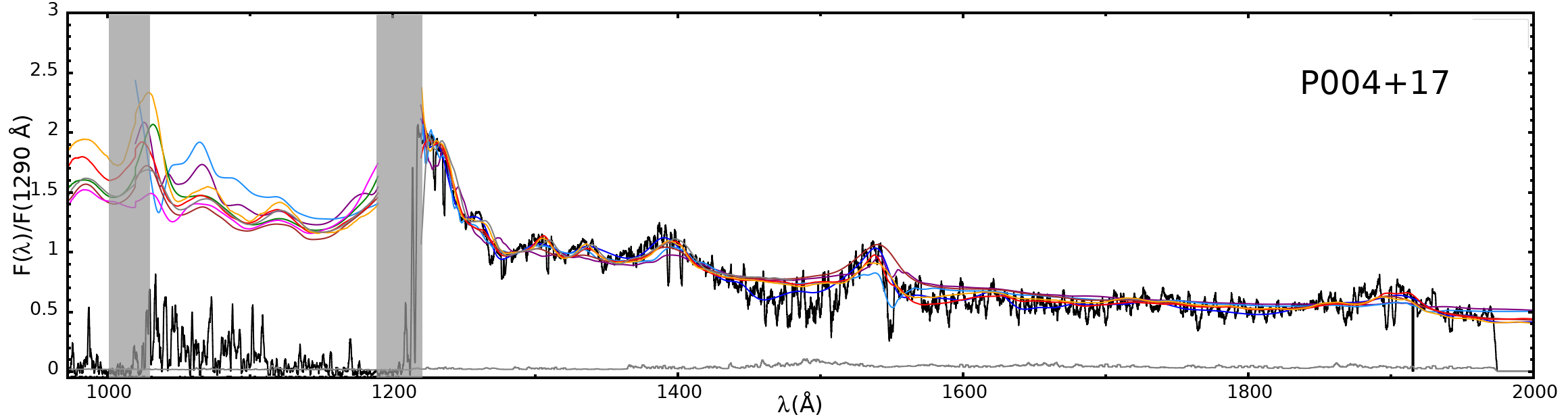}
\includegraphics[width=\textwidth]{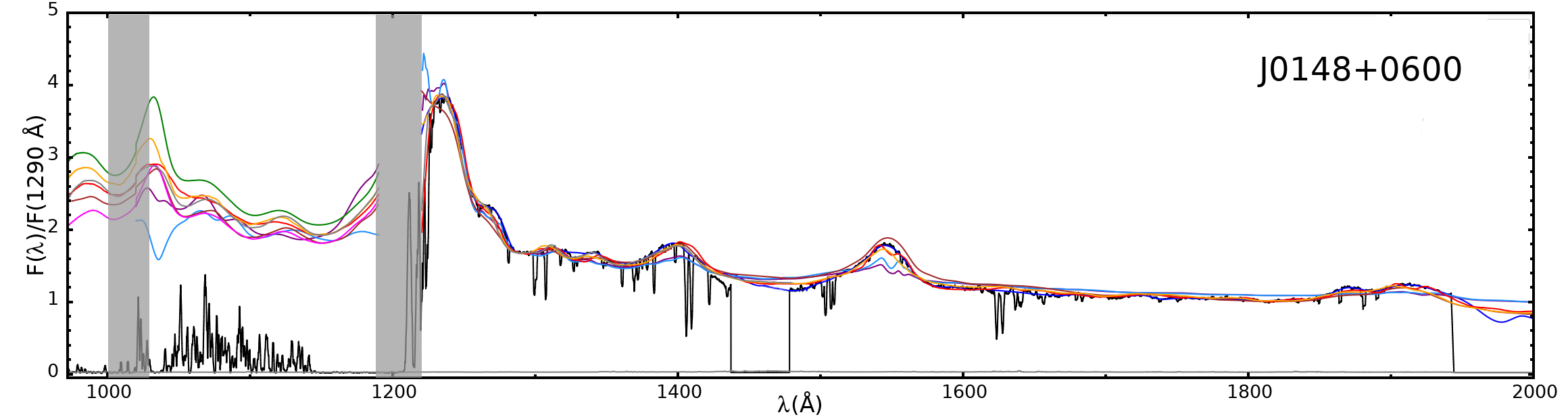}
\includegraphics[width=\textwidth]{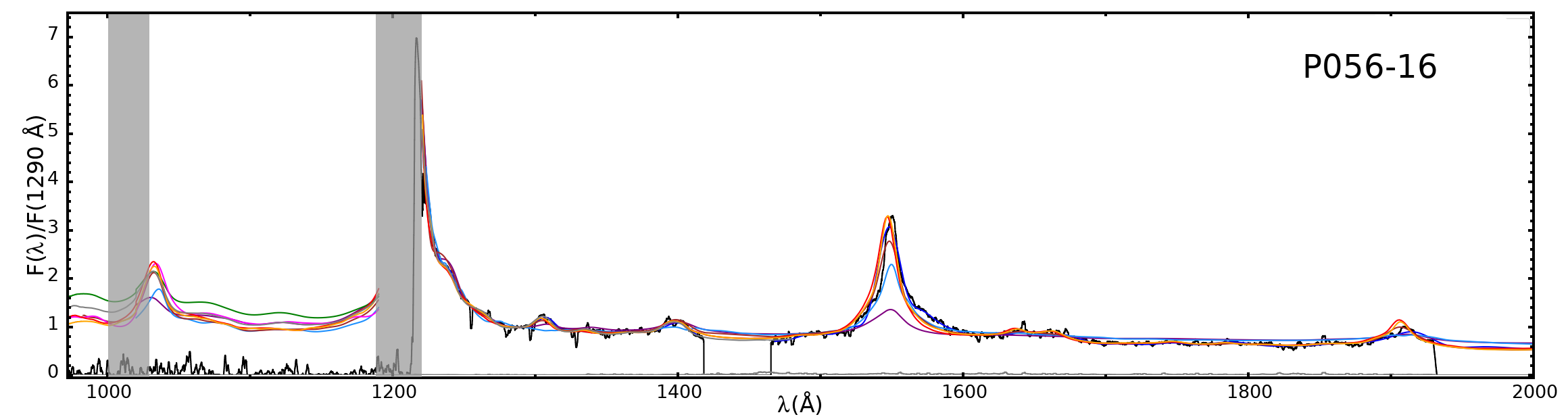}
\end{figure*}
\begin{figure*}
\includegraphics[width=\textwidth]{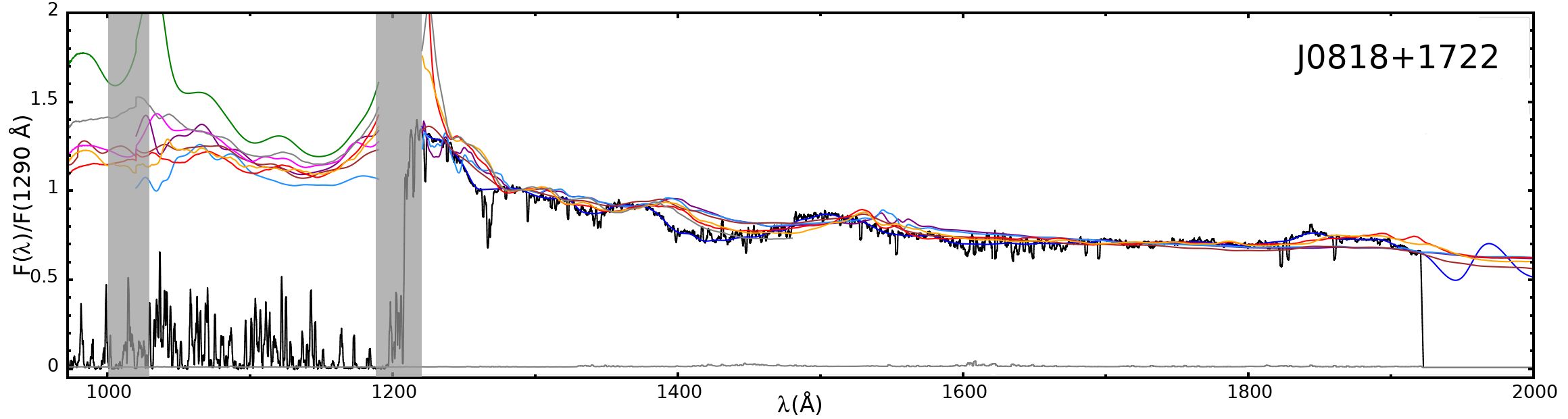}
\includegraphics[width=\textwidth]{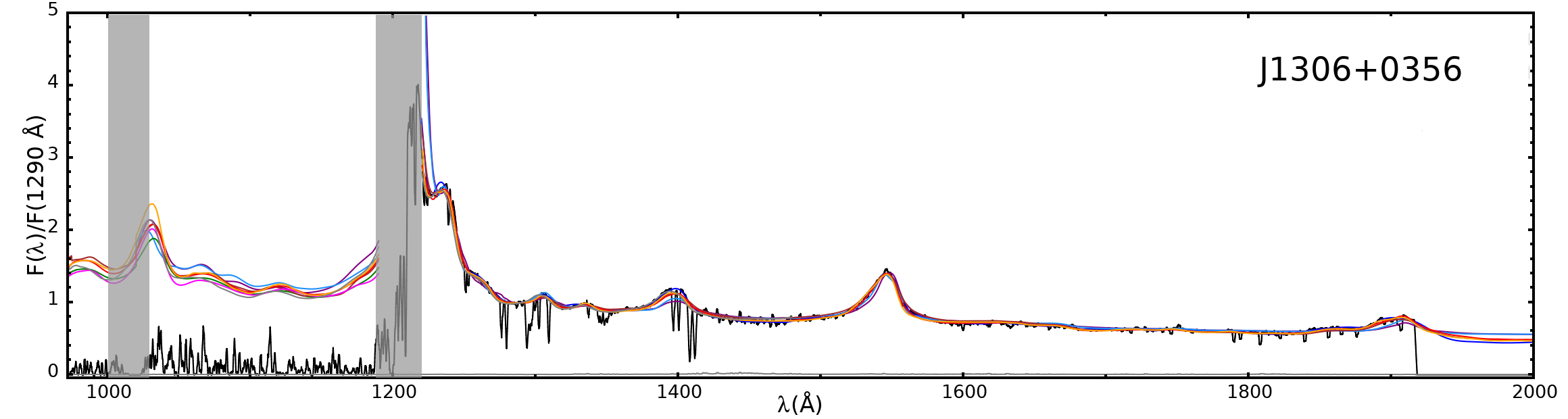}
\includegraphics[width=\textwidth]{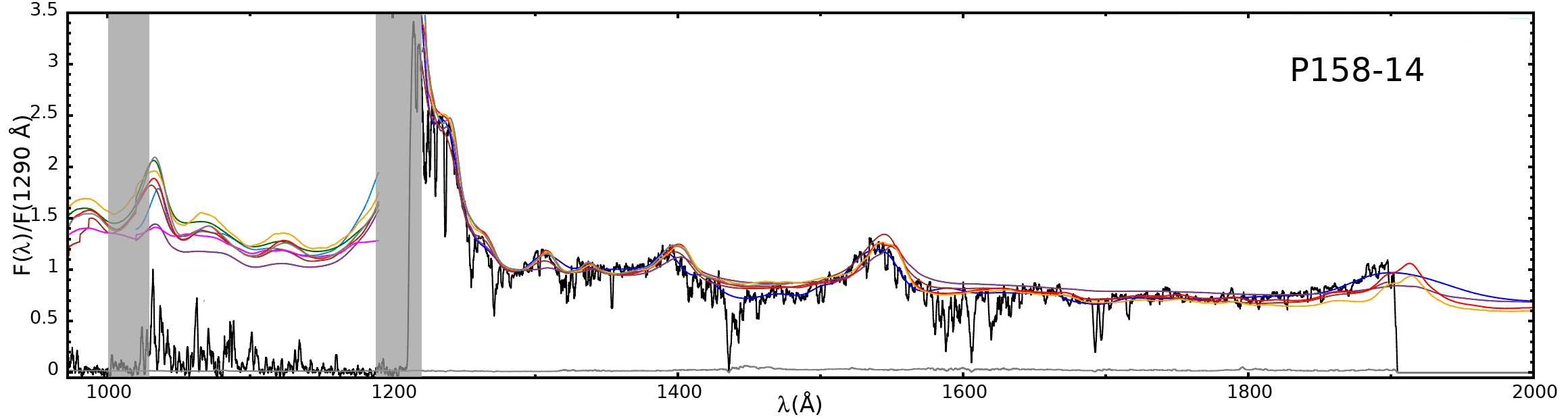}
\includegraphics[width=\textwidth]{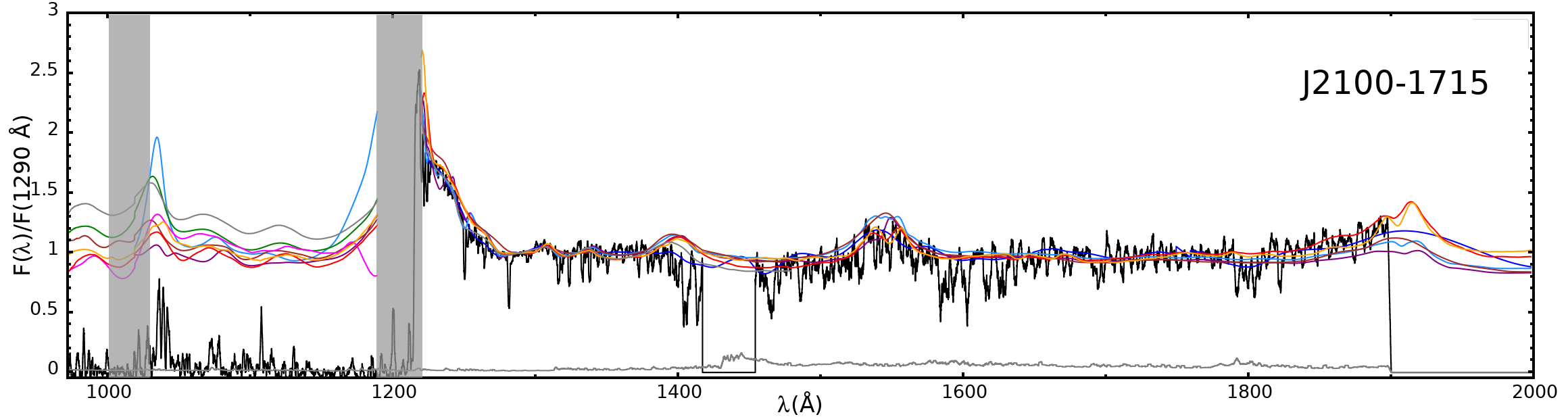}
\includegraphics[width=\textwidth]{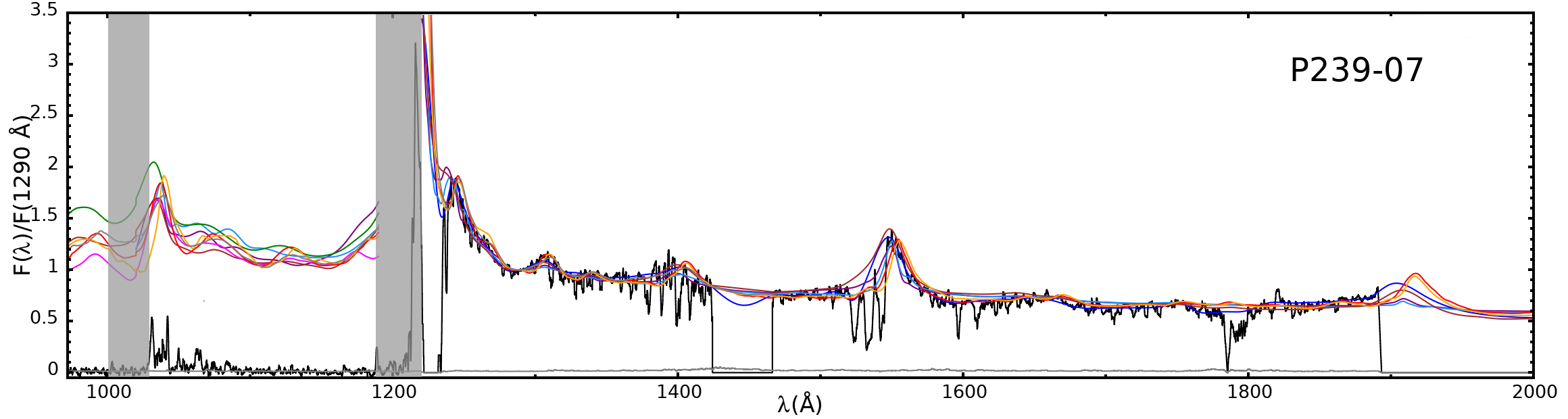}
\end{figure*}
\begin{figure*}
\includegraphics[width=\textwidth]{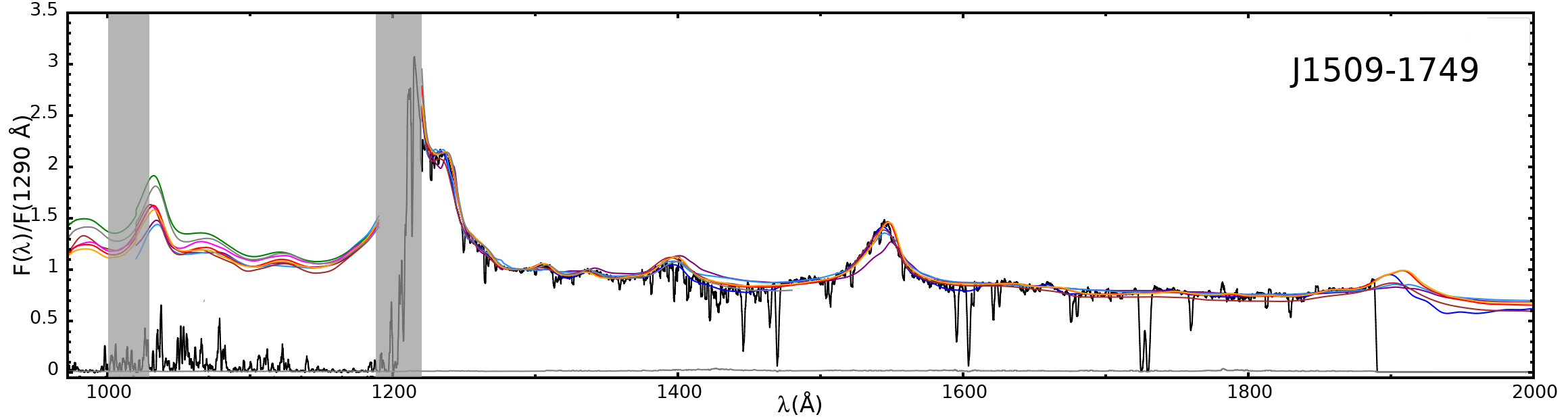}
\includegraphics[width=\textwidth]{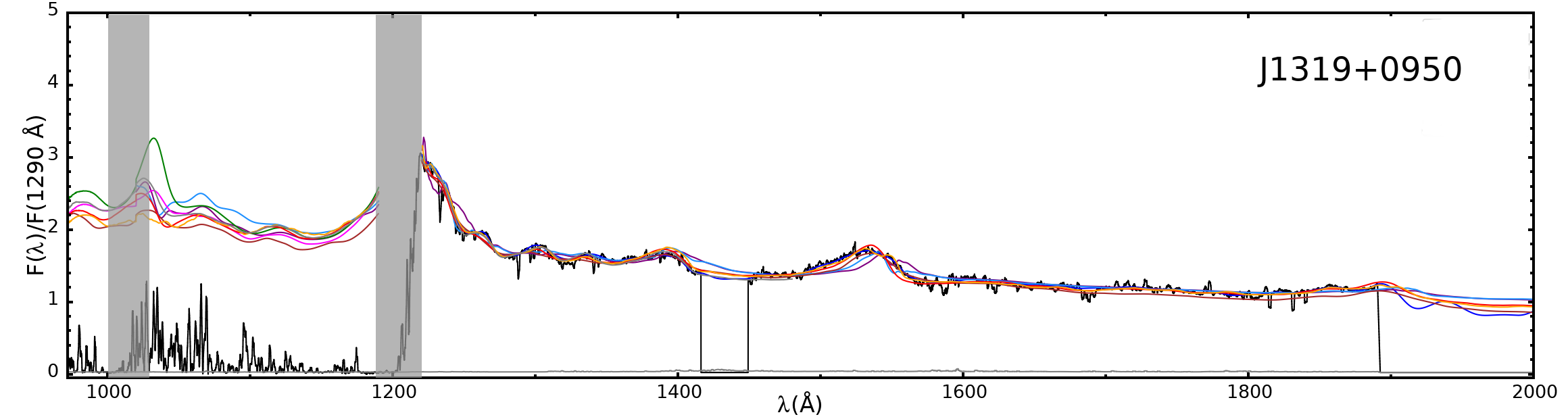}
\includegraphics[width=\textwidth]{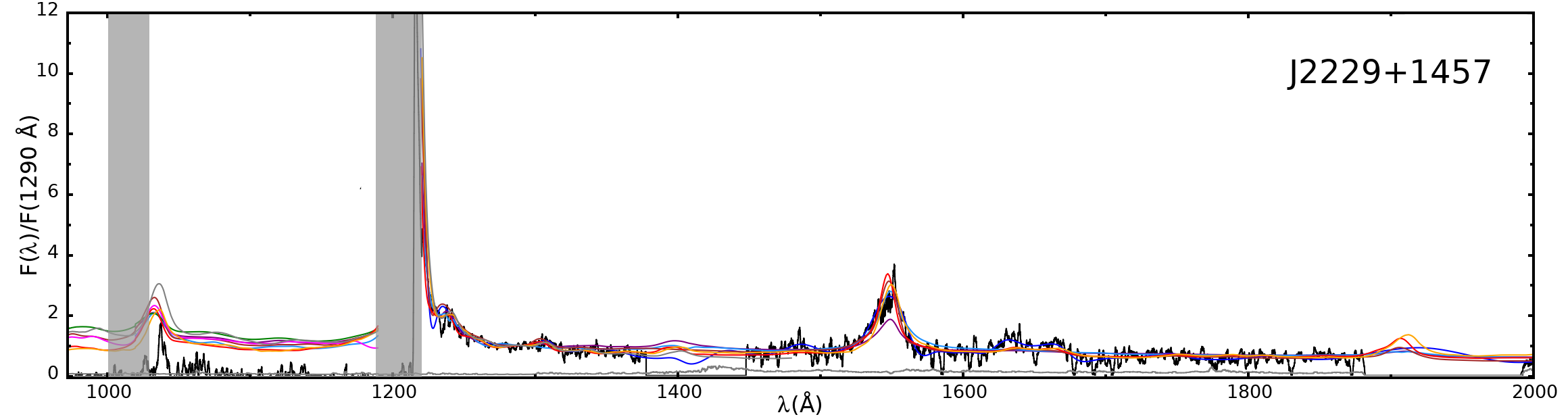}
\includegraphics[width=\textwidth]{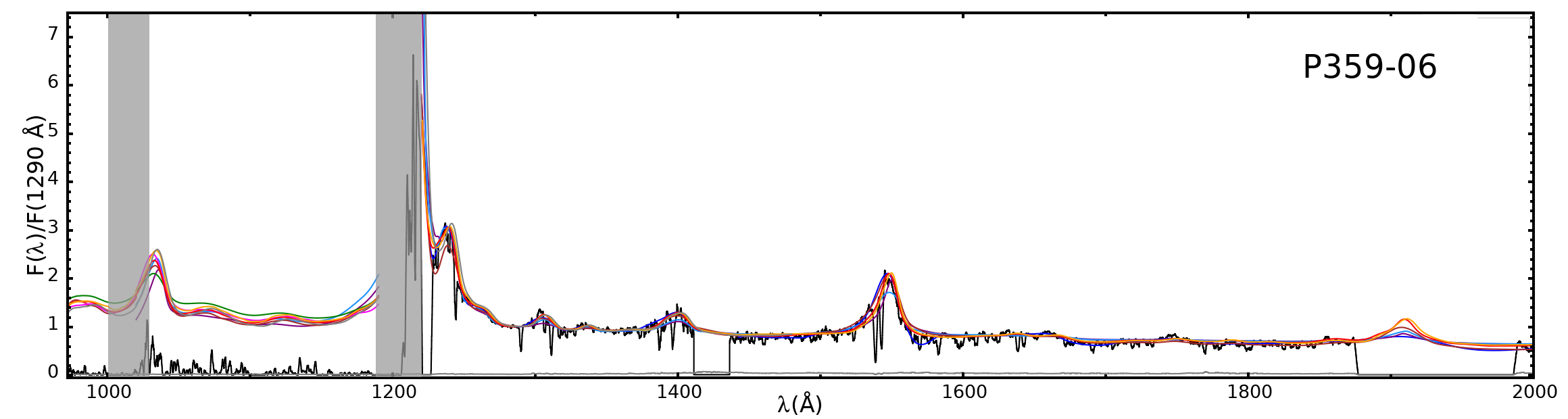}
\includegraphics[width=\textwidth]{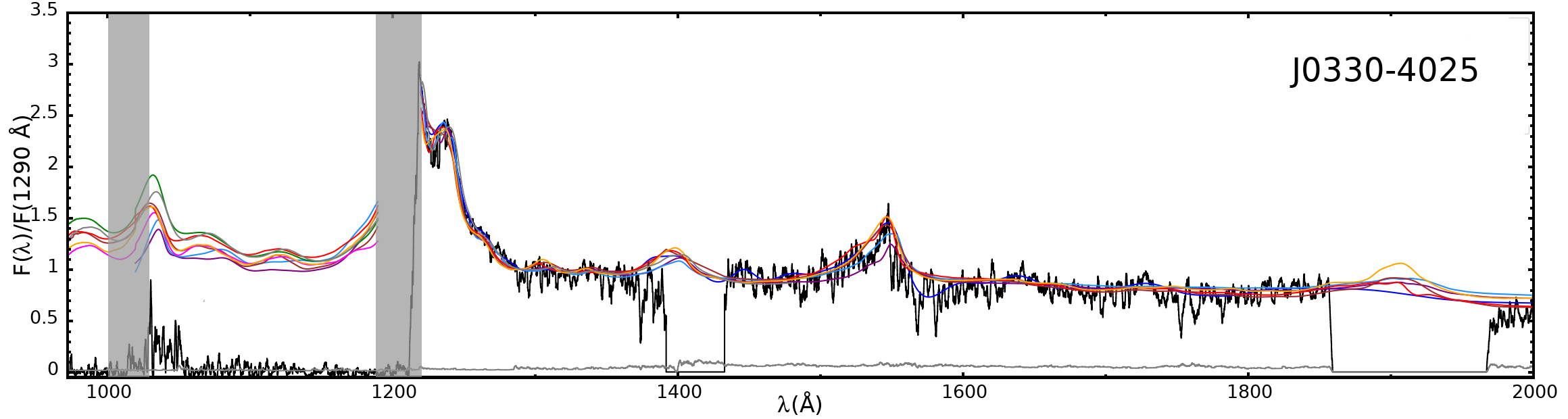}
\end{figure*}

\begin{figure*}
\includegraphics[width=\textwidth]{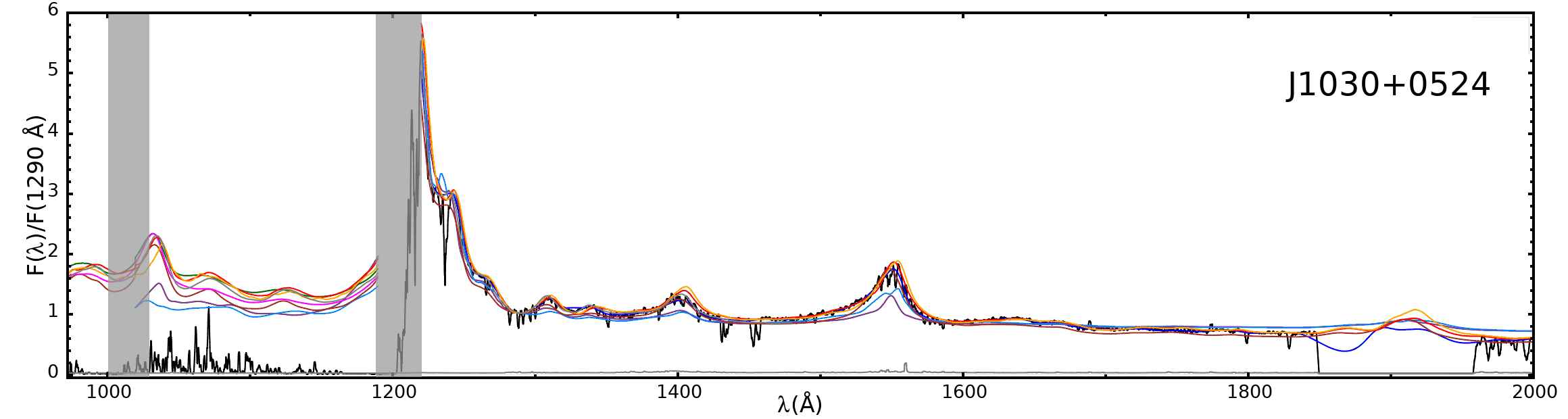}
\includegraphics[width=\textwidth]{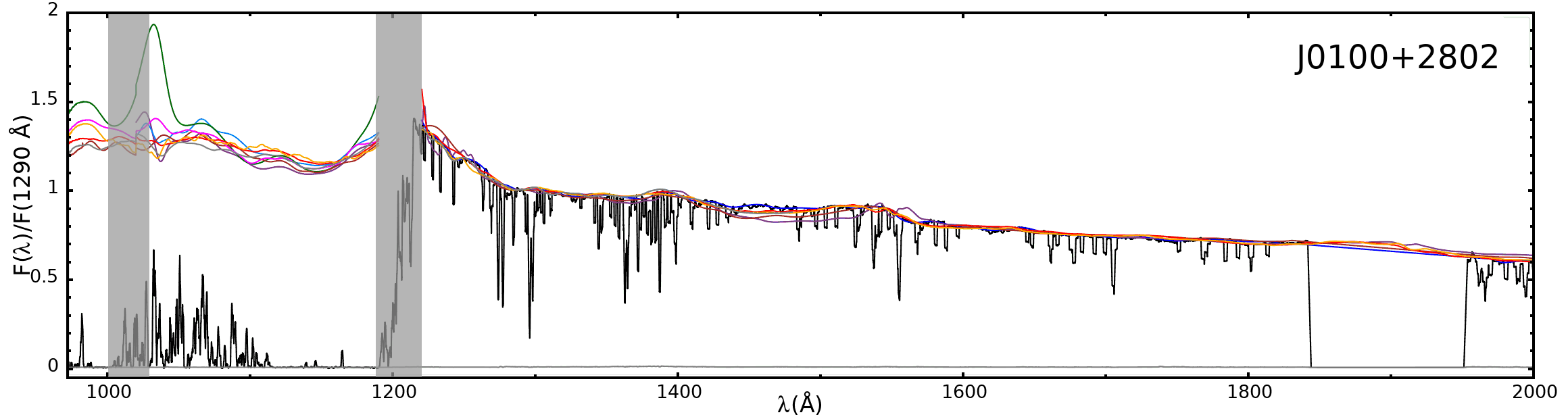}
\includegraphics[width=\textwidth]{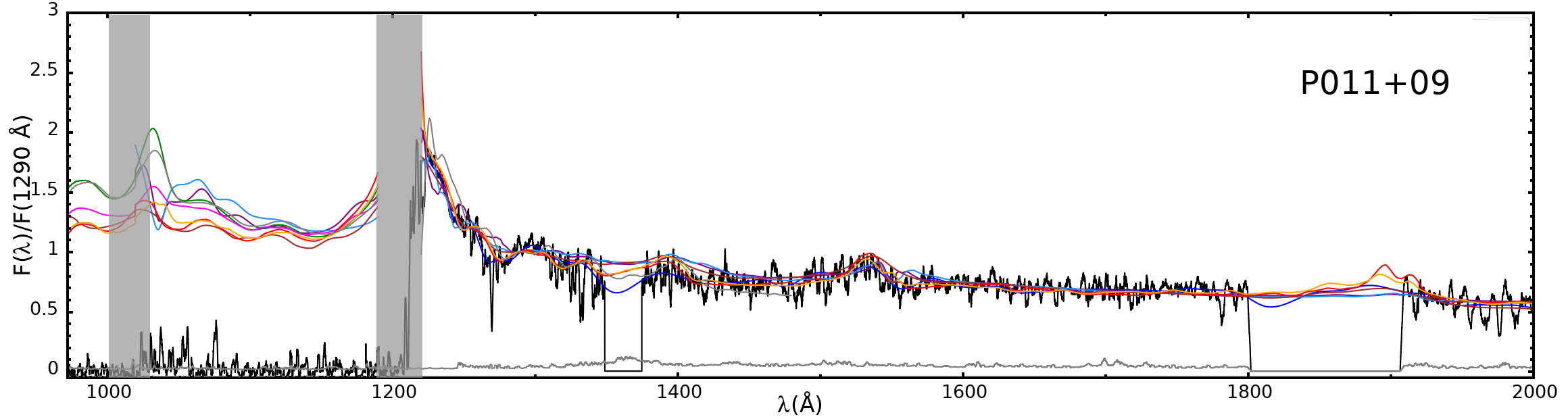}
\includegraphics[width=\textwidth]{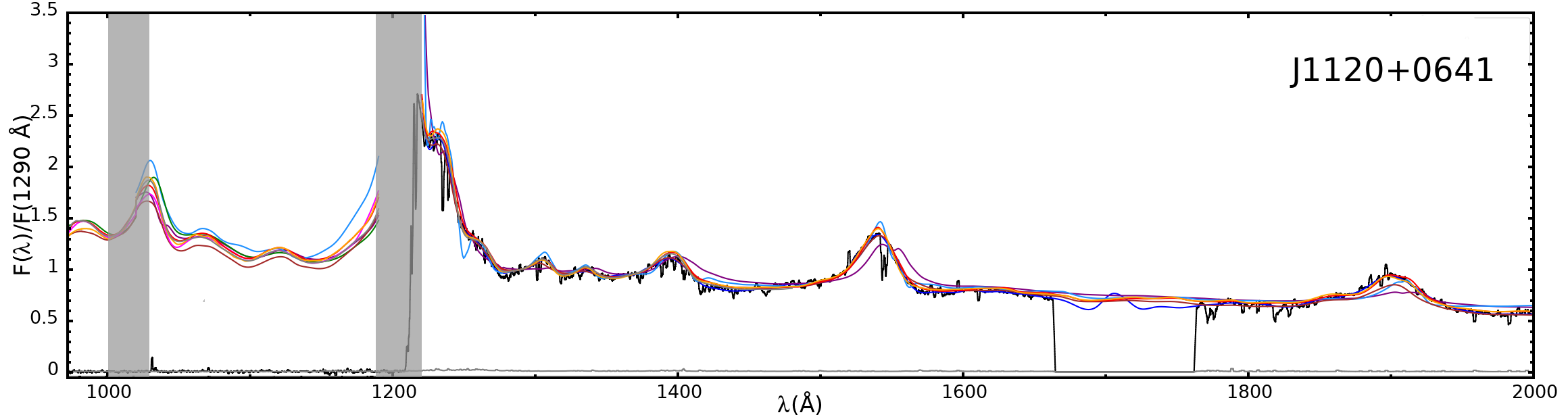}
\end{figure*}


%%%%%%%%%%%%%%%%%%%%%%%%%%%%%%%%%%%%%%%%%%%%%%%%%%

% Don't change these lines
\bsp	% typesetting comment
\label{lastpage}
\end{document}